\documentclass[reprint,onecolumn,amsmath,amssymb,aip,footinbib]{revtex4-2}

\usepackage{graphicx}
\usepackage[colorlinks=true,urlcolor=blue,citecolor=blue,linkcolor=blue]{hyperref}
\usepackage{cleveref}
\usepackage{bm}
\usepackage{newtxtext,newtxmath}
\usepackage{gensymb}
\usepackage{multirow,makecell}
\newcolumntype{C}[1]{>{\centering\arraybackslash}m{#1}}

\newcommand{\rmd}{\mathrm{d}}
\newcommand{\micron}{\mu\mathrm{m}}
\renewcommand{\vec}{\bm}

\newcommand{\abs}[1]{{\left|{#1}\right|}}
\newcommand{\arms}{a_\text{rms}}

\newcommand{\changed}[1]{#1}

\begin{document}

\title{Simulations of laser-driven strong-field QED with \textit{Ptarmigan}:\\
Resolving wavelength-scale interference and $\gamma$-ray polarization}

\author{T. G. Blackburn}
\email{tom.blackburn@physics.gu.se}
\affiliation{Department of Physics, University of Gothenburg, SE-41296 Gothenburg, Sweden}
\author{B. King}
\affiliation{Centre for Mathematical Sciences, University of Plymouth, Plymouth, PL4 8AA, United Kingdom}
\author{S. Tang}
\affiliation{College of Physics and Optoelectronic Engineering, Ocean University of China, Qingdao, Shandong, 266100, China}

\date{\today}

\begin{abstract}
Accurate modelling is necessary to support precision experiments investigating strong-field QED phenomena.
This modelling is particularly challenging in the transition between the perturbative and nonperturbative regimes, where the normalized laser amplitude $a_0$ is comparable to unity and wavelength-scale interference is significant.
Here we describe how to simulate nonlinear Compton scattering, Breit-Wheeler pair creation, and trident pair creation in this regime, using the Monte Carlo particle-tracking code \textit{Ptarmigan}.
This code simulates collisions between high-intensity lasers and beams of electrons or $\gamma$ rays, primarily in the framework of the locally monochromatic approximation (LMA).
We benchmark our simulation results against full QED calculations for pulsed plane waves and show that they are accurate at the level of a few per cent, across the full range of particle energies and laser intensities.
This work extends our previous results to linearly polarized lasers and arbitrarily polarized $\gamma$ rays.
\end{abstract}

\maketitle

\section{Introduction}

Experiments are planned or underway to measure nonlinear Compton and Breit-Wheeler pair-creation in the strong-field QED regime~\cite{dipiazza.rmp.2012,gonoskov.rmp.2022,fedotov.pr.2023}, where the quantum parameter $\chi$ is of order unity.
For plane waves, $\chi$ is the product of the (classical) dimensionless laser amplitude, $a_{0}$, and the (quantum, linear) energy parameter $\eta$.
One promising experimental strategy is to collide a multi-terawatt laser pulse with a beam of ${>}10$-GeV electrons, achieving $a_0 \sim O(1)$ and $\eta \sim O(0.1)$, and thereby reaching the strong-field regime~\cite{bula.prl.1996,burke.prl.1997,meuren.exhilp.2019,abramowicz.epjst.2021,abramowicz.2023}.
In this case, quantum interference has a significant effect, because the formation lengths of all strong-field QED processes can be comparable to the laser wavelength.
The polarization of the $\gamma$ rays, i.e. high-energy photons, that drive electron-positron pair creation also has an important role to play~\cite{king.pra.2013,seipt.njp.2021,dai21}.
Indeed, earlier theory work suggests that $\gamma$-ray polarization has a larger effect on the total probability than fermion spin, as seen when comparing the nonlinear trident process~\cite{King:2013osa,tang.prd.2023} with double nonlinear Compton scattering~\cite{King:2014wfa,seipt.pra.2018}.
Accounting for both interference and polarization effects is necessary for numerical simulations to achieve an accuracy that matches the expected precision of experimental investigations of the transition regime ($a_0 \sim 1$), e.g. LUXE~\cite{abramowicz.epjst.2021,abramowicz.2023}.

In this work we present the \textit{Ptarmigan} simulation framework, which resolves wavelength-scale interference effects by means of the locally monochromatic approximation (LMA)~\cite{bamber.prd.1999,cain,Hartin:2018egj,heinzl.pra.2020,Torgrimsson:2020gws} and how photon emission and electron-positron pair creation depend on the polarization of the high-energy photons.
As we have already demonstrated that LMA-based simulations accurately model these processes in circularly polarized electromagnetic waves~\cite{blackburn.njp.2021,blackburn.epjc.2022}, we consider here the interaction with linearly polarized electromagnetic waves.
This is a richer and more physically relevant problem because of the broken symmetry of linear polarization.
The electric and magnetic fields oscillate along fixed directions, instead of rotating around the propagation axis, thereby defining a preferred direction in space.
As a result, the angular profile of the radiation emitted by electrons travelling through an linearly polarized wave lacks rotational symmetry: it is dipolar (along the $B$-field direction) for small $a_0$, before becoming increasing elliptical (along the $E$-field direction) at larger $a_0$~\cite{harshemesh.ol.2012,seipt.pra.2013}.
Furthermore, the emitted $\gamma$-ray photons are strongly polarized parallel to the laser $E$-field~\cite{milburn.prl.1963,aru.pl.1963,ivanov.epjc.2004,king.pra.2020,Tang:2020xlj}.
Inducing secondary processes with this high-energy radiation provides opportunities to test how strong-field QED processes depend on $\gamma$-ray polarization.
Finally, from a practical perspective, high-power laser systems are naturally linearly polarized to begin with.
Converting to circular polarization at fixed laser energy leads to a peak electric field at focus that is reduced by a factor of $\sqrt{2}$.
It follows that the classical and quantum nonlinearity parameters, $a_0$ and $\chi$, are larger for linear polarization and hence nonlinear and quantum effects are also larger.

We begin by reviewing the probability rates for photon emission (or nonlinear Compton scattering, $e \to e \gamma$) and electron-positron pair creation ($\gamma \to e^+ e^-$) in monochromatic, linearly polarized electromagnetic waves in \cref{sec:Rates}.
We then discuss how the \textit{Ptarmigan} simulation framework combines Monte Carlo sampling of these probability rates with tracking of cycle-averaged classical trajectories to generate predictions for final-state particle spectra, as well as the alternative models that are available, in \cref{sec:Simulations}.
Comparisons with full QED results are presented in \cref{sec:Benchmarking}, which validate the accuracy of the underlying approach.
We also present some examples of the physics that can be explored with \textit{Ptarmigan} in \cref{sec:Examples}.

\section{Probability rates}
\label{sec:Rates}

The probability rates for QED processes in monochromatic electromagnetic waves are controlled by two Lorentz-invariant parameters: the normalized root-mean-square (r.m.s.) amplitude of the wave\footnote{In this section we assume that $\arms$ is a constant. In a simulation framework based on the locally monochromatic approximation, it becomes a slowly varying function of space and time, $\arms(X)$.}, $\arms = e E_\text{rms} / (m \omega)$, and a (quantum) energy parameter $\eta = k.q/m^2$.
Here $e$ is the elementary charge, $m$ is the electron mass, $E_\text{rms}$ the r.m.s. electric field, $k$ is the wavevector of the background, $\omega = |\vec{k}|$ is its frequency, and $q$ is the cycle-averaged momentum (or \emph{quasimomentum}) of the incoming particle.
We set $\hbar$ and $c$ to unity throughout unless otherwise stated.
In a plane EM wave, the quasimomentum of an electron or positron with asymptotic momentum $p$ is $q = p + m^2 \arms^2 k / (2 k.p)$, where $q^2 = m^2 (1 + \arms^2)$.
The peak and r.m.s. amplitudes are related by $a_0 = \sqrt{2} \arms$ for linear polarization and $a_0 = \arms$ for circular polarization.
The quasimomentum of a photon, $q_\gamma$, coincides with the asymptotic momentum, $k'$, as photons are uncharged and massless.

The differential rates depend on the properties of the particle in the initial state through the two parameters $\arms$ and $\eta$.
As is characteristic of QED processes in plane EM waves with a well-defined carrier frequency, they can be written as a sum over an integral harmonic index $n$.
The dependence on the momenta of the daughter particles is parametrized in terms of $s$, the fraction of the parent lightfront momentum transferred to one of the daughters, and the polar and azimuthal angles $\theta$ and $\phi$.
For photon emission, $s = k.k' / k.q$, where $k'$ is the momentum of the emitted photon and $q$ the quasimomentum of the electron or positron.
For pair creation, $s = k.q' / k.k'$, where $q'$ is the quasimomentum of the produced positron and $k'$ the momentum of the decaying photon.
The two angles are conveniently defined in the zero-momentum frame (ZMF) of the scattering event.
The polar angle $\theta$ is determined by kinematics if $s$ and $n$ are known; thus it does not appear explicitly in the theoretical results we will show.
The azimuthal scattering angle $\phi$ is the angle between $\vec{e}$, one of the two vectors defining the background field's polarization, and the projection of the emitted photon (or created positron) momentum $\vec{k}'$ ($\vec{q}$) on the $\vec{e}$-($\vec{e}\times\vec{k}$) plane.

In this work we use rates that are fully resolved in the polarization of the high-energy photons, whether those photons are in the initial or final state.
This ensures that we capture how the positron yield in trident pair creation ($e \to e\gamma \to e e^+e^-$) depends on the polarization of the intermediate photon~\cite{king.pra.2013}.
At the same time, the rates are averaged over the spin of the initial-state electron (or positron) and summed over the spin states of any final-state electrons or positrons.
We make this assumption despite the fact that, at high intensity $a_0 \gg 1$, the radiation power of electrons that are polarized parallel or antiparallel to the magnetic field differs by 30\%; or that as they continue to radiate, electrons spin-polarize antiparallel to the local magnetic field, in an analogy of the Sokolov-Ternov effect~\cite{delsorbo.pra.2017}.
This is because we consider the interaction with \emph{plane-wave-like} fields, where the electric and magnetic fields change sign every half-cycle (linear polarization) or rotate around the propagation axis (circular polarization).
Thus initially unpolarized electrons do not accumulate significant polarization over the course of the interaction, particularly if the laser pulse is more than a few cycles in duration~\cite{seipt.pra.2018}.
There are interaction scenarios where this does occur, but generally some symmetry-breaking mechanism is required, such as a laser field with two colours~\cite{seipt.pra.2019} or a small degree of elliptical polarization~\cite{li.prl.2019} (the latter combined with an angular cut).
Here it is sufficient to treat electrons and positrons as always being unpolarized; we will return to this point in future work.

The polarization of the emitted (or decaying) photon is defined in terms of the three Stokes parameters $S_1$, $S_2$ and $S_3$.
The two basis vectors used in deriving the theoretical results are\footnote{%
These are equivalent to the manifestly covariant basis vectors $\varepsilon_{1,2} = \epsilon_{1,2} - \delta_{1,2} k$, where $\delta_{1,2} = (\epsilon_{1,2} \cdot k') / (k \cdot k')$, $\epsilon_1^\mu = (0,1,0,0)$, $\epsilon_2^\mu = (0,0,1,0)$ and $k^\mu = \omega (1,0,0,1)$~\cite{king.pra.2020}, which are orthonormal in the four-dimensional sense: $\varepsilon_1 \cdot \varepsilon_2 = 0$, $\varepsilon_{1,2} \cdot k' = 0$ and $\varepsilon_{j}\cdot\varepsilon_{j}=-1$ for $j\in\{1,2\}$.
We make these orthonormal in the three-dimensional sense, at the expense of explicit covariance, by transforming $\varepsilon_{1,2} \to \varepsilon_{1,2} + \delta_{1,2} \omega k' / \omega'$.}
    \begin{align}
    \vec{\varepsilon}_1 &= \vec{\hat{E}} + \frac{k'_x}{\omega' - k'_z} \left( \vec{n} - \vec{n}' \right),
    &
    \vec{\varepsilon}_2 &= \vec{\hat{B}} + \frac{k'_y}{\omega' - k'_z} \left( \vec{n} - \vec{n}' \right),
    \label{eq:LMAbasis}
    \end{align}
where $\vec{n}$ and $\vec{n}'$ are unit vectors along the laser wavevector ($\vec{k}$) and photon momentum ($\vec{k}'$), respectively.
The basis vectors $\vec{\varepsilon}_1$ and $\vec{\varepsilon}_2$ form an orthonormal triad with $\vec{n}' = \vec{k}' / \omega'$.
Here $\omega' = \abs{\vec{k}'}$ and the coordinate system is defined by unit vectors parallel to the laser electric field ($\vec{\hat{E}}$), magnetic field ($\vec{\hat{B}}$) and wavevector.
$S_1$, $S_2$ and $S_3$ are: the degree of linear polarization with respect to the basis given in \cref{eq:LMAbasis}; the degree of linear polarization with respect to the same basis, but rotated by $45\degree$ around $\vec{k}'$; and the degree of circular polarization.
We will refer to photons with $S_1 = -1$, which are polarized parallel to the laser electric field, as ``$E$-polarized'', and to photons with $S_1 = +1$, which are polarized parallel to the laser magnetic field, as ``$B$-polarized''.

In what follows, we give the double-differential emission rates per unit proper time in azimuthal angle $\phi$ and lightfront momentum fraction $s$ for the nonlinear Compton and Breit-Wheeler processes.
If integrated over a plane-wave background, they are related to the probability $\textsf{P}$ of the process via:
    \begin{align}
    \textsf{P} &= \int \! \rmd\tau \, W,
    \label{eqn:P1}
    \\
    W &= \sum_{n = n^\star}^{\infty}
        \int_{s_n^-}^{s_n^+} \! \rmd s
        \int_0^{2\pi} \! \rmd \phi
            \, \frac{\rmd^2 W_n}{\rmd s \rmd \phi}
    \end{align}
where $n^{\star}$ is the threshold harmonic, the lightfront momentum fraction limits $s_{n}^{-}$ and $s_{n}^{+}$ can depend on harmonic order, and $\tau$ is the proper time.
Strictly, \cref{eqn:P1} is valid only when the integral is small compared to one;
it may be interpreted as the limiting case of $\textsf{P} = 1 - \exp(-{\int\!\rmd\tau \,W})$, the probability for at least one event to occur.
We focus here on the case that the background field is a linearly polarized electromagnetic wave; for completeness, the equivalent results for a circularly polarized wave and a constant, crossed field are presented in \cref{sec:CPwave,sec:CCF} respectively.

\subsection{Photon emission}

Emission of a single, high-energy photon is often called \emph{nonlinear Compton scattering} in the context of laser interactions.
The double-differential emission rate per unit proper time, $W^\gamma$, at a particular harmonic index $n$, in the LMA is given by~\cite{heinzl.pra.2020} (see also \cite{nikishov.jetp.1964}):
    \begin{equation}
    \frac{\rmd^2 W^\gamma_n}{\rmd s \rmd\phi} =
        -\frac{\alpha m}{2 \pi}
            \left\{
            A_0^2(n, x, y)
            + \arms^2 \left[ 2 + \frac{s^2}{1-s} \right]
            \left[ A_0(n, x, y) A_2(n, x, y) - A_1^2(n, x, y) \right]
            \right\}
    \label{eq:LPPhotonRate}
    \end{equation}
where the arguments of the $A$ functions (defined in \cref{sec:DoubleBessel}) are
    \begin{align}
    x &= -2 n \cos\phi \sqrt{\frac{2 \arms^2 w_n (1 - w_n)}{1 + \arms^2}},
    &
    y &= \frac{\arms^2}{4 \eta_e} \frac{s}{1-s},
    &
    w_n &= \frac{s}{s_n (1 - s)},
    &
    s_n &= \frac{2 n \eta_e}{1 + \arms^2},
    \end{align}
and $0 \leq s \leq s_n / (1 + s_n)$.
The energy parameter of the electron $\eta_e = k.q / m^2$.
The total rate $W^\gamma$ is obtained by integrating \cref{eq:LPPhotonRate} over $s$ and $\phi$ and summing the resulting partial rates $W^\gamma_n$ over all $n \geq 1$.

The Stokes parameters of the emitted photon are given by~\cite{li.jjap.2003,ivanov.epjc.2004}:
    \begin{align}
    \begin{split}
    S_1 &= \frac{1}{S_0} \left[ 2(A_1^2 - A_0 A_2) - (1 + 2 r_n^2 \sin^2\phi) \frac{A_0^2}{\arms^2} \right],
    \\
    S_2 &= \frac{1}{S_0} \left [\frac{r_n^2 A_0^2}{\arms^2} \sin 2\phi + \frac{4 r_n A_0 A_1}{\sqrt{2} \arms} \sin\phi \right],
    \\
    S_3 &= 0,
    \end{split}
    \label{eq:LPNLCStokes}
    \end{align}
where
    \begin{align}
    S_0 &= \left( 1 - s + \frac{1}{1-s} \right) ( A_1^2 - A_0 A_2 ) - \frac{A_0^2}{\arms^2},
    &
    r_n^2 &= \frac{2 n \eta_e (1-s)}{s} - (1 + \arms^2).
    \end{align}
These are defined with respect to the basis given in \cref{eq:LMAbasis}.
In the limit $s \to 0$, where emission is dominated by the first harmonic $n = 1$, $S_1 = \cos 4\phi$ and $S_2 = \sin 4\phi$.
In general, the radiation is only partially polarized.

\paragraph{Classical limit (nonlinear Thomson scattering).}

In the limit that $\eta_e \to 0$, $x$ and $y$ have universal shapes as functions of $v = s / s_n$.
    \begin{equation}
    \frac{\rmd^2 W^{\gamma,\text{cl}}_n}{\rmd v \rmd\phi} =
        -\frac{\alpha m s_n}{2 \pi}
            \left\{
            A_0^2(n, x, y)
            + 2 \arms^2
            \left[ A_0(n, x, y) A_2(n, x, y) - A_1^2(n, x, y) \right]
            \right\}.
    \label{eq:LPClassicalPhotonRate}
    \end{equation}
The arguments of the $A$ functions are
    \begin{align}
    x &= -2 n \cos\phi \sqrt{\frac{2 \arms^2 v (1 - v)}{1 + \arms^2}},
    &
    y &= \frac{n \arms^2 v}{2 (1 + \arms^2)},
    \end{align}
and $0 \leq v \leq 1$.
The total rate is proportional to $\eta_e$.
The Stokes parameters are obtained by using $r_n^2 = (1 + \arms^2) (1-v)/v$ and replacing $1 - s + 1/(1-s) \to 2$ in \cref{eq:LPNLCStokes}.

\paragraph{Low-intensity limit (linear Compton scattering).}

The arguments of the Bessel functions $x$ and $y$ reach their largest values when $s = s_n / (2 + s_n)$ and $\theta = 0$, where $x = \sqrt{2} n \arms / \sqrt{1 + \arms^2}$ and $y = n \arms^2 / [4 (1 + \arms^2)]$.
In the linear regime, $\arms \to 0$, we may therefore expand $J_n(x,y)$ around $x = y = 0$.
As $x$ is linear in $\arms$, and $y$ quadratic, this expansion is done to order $x^{2k}$ and $y^k$.

For the first harmonic, we find (expanding to $k = 1$):
    \begin{equation}
    \frac{\rmd^2 W^\gamma_1}{\rmd s \rmd\phi} =
        \frac{\alpha m \arms^2}{8 \pi}
        \left\{
            1 - s + \frac{1}{1-s}
            + \left[ \frac{2 s^2}{\eta_e^2 (1 - s)^2} - \frac{4 s}{\eta_e(1-s)}  \right] \cos^2\phi
        \right\},
    \end{equation}
where $0 \leq s \leq 2\eta_e / (1 + 2 \eta_e)$, and
    \begin{equation}
    W^\gamma_1 = \frac{\alpha m \arms^2}{4} \left[
            \frac{1}{2} + \frac{4}{\eta_e} - \frac{1}{2 (1+2\eta_e)^2}
            + \left( 1 - \frac{2}{\eta_e} - \frac{2}{\eta_e^2} \right) \ln(1+2\eta_e)
        \right].
    \end{equation}
At the first Compton edge $s = 2\eta_e/(1+2\eta_e)$, the Stokes parameters take the limiting values $S_1 = \left(1 + \frac{2\eta_e^2}{1+2\eta_e}\right)^{-2} \leq 1$ and $S_2 = S_3 = 0$.
\changed{The maximum attainable polarization of the radiation is limited by recoil effects, which generate motion in the $1/\gamma$ cone; for $\eta_{e}\ll1$, corrections to $S_{1} = 1$ are proportional to $\hbar$ and powers thereof.}

\subsection{Pair creation}

Production of an electron-positron pair is often called the \emph{nonlinear Breit-Wheeler process} in the context of laser interactions.
The double-differential pair-creation rate per unit ``proper'' time, $W^\pm$, at a particular harmonic index $n$, is given by~\cite{tang.prd.2022} (see also \citet{nikishov.jetp.1964})
    \begin{multline}
    \frac{\rmd^2 W^\pm_n}{\rmd s \rmd\phi} =
        \frac{\alpha m}{2 \pi}
            \left\{
            A_0^2
            + \arms^2 \left[ \frac{1}{s (1-s)} - 2\right]
            \left( A_1^2 - A_0 A_2 \right)
            \right. \\
            - S_1 \left[ (A_0 \, r_n \cos\phi - \sqrt{2} \arms A_1)^2 - (A_0 \, r_n \sin\phi)^2 \right]
            \\ \left.
            - S_2 \left[ 2 A_0 \, r_n \sin\phi \left(A_0 \, r_n \cos\phi - \sqrt{2} \arms A_1 \right) \right]
            \right\}
    \label{eq:PairCreationRate}
    \end{multline}
where the arguments of the functions $A_i \equiv A_i(n, x, y)$ are
    \begin{align}
    x &= \frac{\sqrt{2} \arms r_n \cos\phi}{\eta_\gamma s (1 - s)},
    &
    y &= \frac{\arms^2}{4 \eta_\gamma s (1 - s)},
    &
    r^2_n &= 2 n \eta_\gamma s (1-s) - (1 + \arms^2),
    \end{align}
and the range of $s$ is bound by $|s - 1/2| \leq \sqrt{1/4 - 1/s_n}$.
The photon energy parameter $\eta_\gamma = k.k' / m^2$.
The Stokes parameters appearing in this expression are defined with respect to the basis given in \cref{eq:LMAbasis}.
The total rate is obtained by integrating \cref{eq:PairCreationRate} over all $s$ and $\phi$, then summing over all harmonic orders $n \geq n^\star = \lceil 2(1+\arms^2)/\eta_\gamma \rceil$.
It depends on the normalised amplitude $\arms$, energy parameter $\eta_\gamma$ and only the first Stokes parameter $S_1$, i.e. $W^\pm = W^\pm(a, \eta_\gamma, S_1)$.

\subsection{Double Bessel functions}
\label{sec:DoubleBessel}

The $A$ functions are defined in terms of the `double Bessel functions' $J_n(x,y)$:
    \begin{align}
    \begin{split}
    A_0(n, x, y) &= J_n(x, y),
    \\
    A_1(n, x, y) &= \frac{J_{n-1}(x, y) + J_{n+1}(x, y)}{2},
    \\
    A_2(n, x, y) &= \frac{J_{n-2}(x, y) + 2J_n(x, y) + J_{n+2}(x, y)}{4}.
    \end{split}
    \end{align}
The double Bessel functions extend the usual Bessel functions $J_n(x)$ in the following way:
    \begin{equation}
    J_n(x, y) = \frac{1}{2 \pi} \int_{-\pi}^{\pi} \exp[- i n \theta + i x \sin\theta - i y \sin 2\theta] \, d\theta = \sum_{r = -\infty}^{\infty} J_{n+2r}(x) J_n(y).
    \end{equation}
The $A$ functions satisfy the following equality, when evaluated at the same $n, x, y$:
    \begin{equation}
    (n - 2 y) A_0 - x A_1 + 4 y A_2 = 0.
    \end{equation}
We have implemented the recurrence algorithm introduced by \citet{lotstedt.pre.2009} in order to evaluate these functions with the necessary speed and accuracy.
The kinematics of Compton scattering and pair creation mean that our implementation needs to consider only the domain $0 \leq x < n \sqrt{2}$ and $0 \leq y < n/2$ at fixed $n$.

\section{Simulation framework}
\label{sec:Simulations}

By default, the \textit{Ptarmigan} simulation code models particle and photon dynamics in QED by means of the LMA.
This mode will be discussed in detail here and will be the focus of the benchmarking presented in \cref{sec:Benchmarking}.
However, the code also includes modes based on classical electrodynamics and the locally constant field approximation (LCFA).
\textit{Ptarmigan}'s coverage of polarization dependence in the two basic strong-field QED processes, as well as the theoretical models by which they are implemented, is summarized in \cref{tbl:Coverage}.
The LMA is available in the parameter region $\eta_{e,\gamma} \leq 1$ and $a_0 \leq 20$, whereas the LCFA can be used for arbitrary values of the same.
In both cases, the laser may be linearly (LP) or circularly polarized (CP).

    \begin{table}[h]
    \begin{tabular}{|C{1.5cm}|C{2.5cm} C{1.5cm} C{1.5cm}|C{2cm} C{2cm} C{2cm}|}
    \hline
    \multirow{2}{*}{Process} & \multicolumn{3}{c|}{Polarization} & \multicolumn{3}{c|}{Available modes} \\
    & $e^\pm$ & $\gamma$ & laser & QED & classical & mod. clas. \\
    \hline
    $e^\pm \to e^\pm \gamma$ & \makecell{averaged (initial),\\summed (final)} & arbitrary & LP / CP & LMA / LCFA & LMA / LCFA & LCFA \\
    $\gamma \to e^+ e^-$ & summed & arbitrary & LP / CP & LMA / LCFA & n/a & n/a \\
    \hline
    \end{tabular}
    \caption{Polarization dependence and available models of photon emission and electron-positron pair creation in \textit{Ptarmigan}.}
    \label{tbl:Coverage}
    \end{table}

\subsection{Default mode}

In the locally monochromatic approximation, the background electromagnetic field is modelled as a wave with a cycle-averaged amplitude $\arms$ and wavevector $k$ that vary slowly in space and time.
Charged particles move through this field on classical trajectories defined by the quasimomentum $q$ and cycle-averaged position $X$, i.e. the slowly varying component of the worldline.
These satisfy the following equations of motion~\cite{quesnel.pre.1998}
    \begin{align}
    \frac{d q_\mu}{d \tau} &= \frac{1}{2} m \partial_\mu \arms^2(X),
    &
    \frac{d X^\mu}{d \tau} &= \frac{q^\mu}{m},
    \label{eq:EOM}
    \end{align}
where $m$ is the electron mass and $\tau$ is the proper time.\footnote{\changed{\textit{Ptarmigan} uses proper time, rather than phase, to parameterize the trajectory of a particle. For context, the} proper time is related to phase $\varphi$ by the energy parameter: $\frac{d \varphi}{d \tau} = m \eta_e$.}
The code solves \cref{eq:EOM} numerically by means of a leapfrog method, thereby tracking electrons and positrons as they travel through the laser pulse.
\changed{Given the particle position and momentum at a particular proper time, $X(\tau)$ and $q(\tau)$, the local energy parameter follows from $\eta_e = k.q / m^2$ and the intensity parameter from $a_\text{rms} = \sqrt{q^2  / m^2 - 1}$.}
Each particle has a \emph{weight}, which determines the statistical importance of that particular particle.
When a simulation is used to model a real experiment, and true particle counts are required as output, the weight is used to indicate the number of physical particles represented by that individual particle.

\textit{Ptarmigan} simulates interactions with plane-wave or focussed laser pulses by taking the normalized squared potential to be~\cite{blackburn.njp.2021}:
    \begin{align}
    \arms^2(X) &= \frac{[a_0 \, g(\varphi)]^2}{1 + (z/z_R)^2}
        \exp\!\left[ -\frac{2 \vec{r}_\perp^2}{w_0^2 + (\vartheta z)^2} \right]
        \times \begin{cases}
            1/2 & \text{LP} \\
            1 & \text{CP}
        \end{cases},
    \label{eq:FocussedPulse}
    \end{align}
where $X^\mu = (t, \vec{r}_\perp, z)$, $w_0$ is the beam waist (the radius at which the intensity falls to $1/e^2$ of its central value), $z_R = \pi w_0^2 / \lambda$ is the Rayleigh range, $\vartheta = w_0 / z_R$ is the diffraction angle, and the pulse envelope $g(\varphi)$ is a function of phase $\varphi = \omega (t - z)$.

\subsubsection{Photon emission}

\begin{figure}
    \centering
    \includegraphics[width=0.4\linewidth]{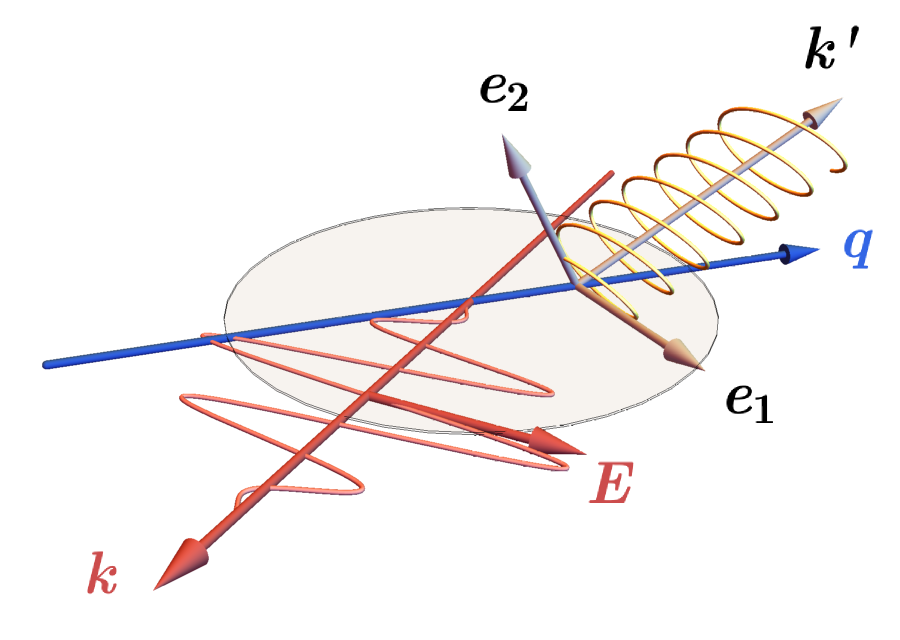}
    \caption{%
        In \textit{Ptarmigan}, the polarization of a high-energy photon (in yellow) emitted by an electron (with quasimomentum $q$, in blue) is defined with respect to the basis illustrated here.
        The first basis vector lies in the plane defined by the laser polarization $\vec{E}$ and wavevector $\vec{k}$ (illustrated by the grey circle), and is perpendicular to the photon momentum $\vec{k}'$.
        The second basis vector is perpendicular to the first and to $\vec{k}'$.
        The Stokes parameters are transformed to a local basis at each timestep so that the pair-creation probability can be evaluated.
        Under the locally monochromatic approximation (LMA), this local basis is defined by \cref{eq:LMAbasis}; under the locally constant field approximation (LCFA) it is defined by the instantanous acceleration (see \cref{sec:CCF}).
        }
    \label{fig:Basis}
    \end{figure}

At each timestep of size $\Delta t$, the probability of photon emission $P^\gamma = W^\gamma(\arms, \eta_e) \Delta t / (q^0 / m)$ is evaluated and a photon generated if $r_1 < P^\gamma$, where $r_1$ is a pseudorandomly generated number in the unit interval.
The total rate of emission $W(\arms, \eta_e)$ is precalculated and tabulated as a function of $\arms$ and $\eta_e$.
We do this by numerically integrating the double-differential rate, \cref{eq:LPPhotonRate} (LP) or \cref{eq:CP_EmissionRate} (CP), over all $s$ and $\phi$ then summing the resulting partial rates all over harmonic orders $1 \leq n \leq n_\text{max}$.
The cutoff $n_\text{max}$ is automatically determined to ensure convergence.
On emission, the harmonic index $n$ is sampled by inverting $r_2 = \text{cdf}(n)$.
Here $r_2$ is another pseudorandom number and $\text{cdf}(n) = \sum_{i=1}^n W^\gamma_i / W^\gamma$ is the cumulative density function, which is also precalculated and tabulated as a function of $\arms$, $\eta_e$ and $n$.
Once the harmonic index has been determined, the lightfront momentum transfer fraction $s$ and azimuthal angle $\phi$ are obtained by rejection sampling of the double-differential rate.
In the ZMF, the photon momentum and polar scattering angle are given by
    \begin{align}
    \abs{\vec{k}_\text{ZMF}'} &= \frac{m n \eta_e}{\sqrt{1 + \arms^2 + 2 n \eta_e}},
    &
    \cos\theta_\text{ZMF} &= 1 - \frac{s (1 + \arms^2 + 2n \eta_e)}{n \eta_e}.
    \end{align}
These quantities, with $\phi$, allow us to construct the photon's four-momentum $k'$, which is then transformed back to the laboratory frame.
The four-velocity of the ZMF with respect to the laboratory frame is $U = (q + n k) / |q + n k|$.
The electron (positron) quasimomentum after the scattering, $q'$, is fixed by momentum conservation
    \begin{equation}
    q + n k = q' + k'.
    \label{eq:PhotonRecoil}
    \end{equation}
The weight of the photon is identical to the weight of the emitting electron.
We then assign the newly emitted photon a set of Stokes parameters, using \cref{eq:LPNLCStokes} (LP) or \cref{eq:CPNLCStokes} (CP).
Note that we do not perform an additional sampling step to project the photon polarization onto a particular eigenstate (as is done in \citet{li.prl.2020}, for example).
In fact, no projection takes place unless the photon reaches a synthetic detector, at which point we are free to choose an arbitrary basis.
Once the Stokes parameters are chosen, they are transformed such that they are defined with respect to a global basis, which is illustrated in \cref{fig:Basis}.
\changed{The code uses this global basis to unify the tracking procedures under the LMA and LCFA, which otherwise define different preferred bases.
All output is defined with respect to the global basis.}

\subsubsection{Pair creation}

Photons, whether externally injected or emitted during the interaction, are tracked by the simulation as they travel through the electromagnetic field on ballistic trajectories.
In the current version of \textit{Ptarmigan}, the only physical process that affects photon propagation is decay to an electron-positron pair.
As each photon represents many real particles, this means that the (complex) polarization vector, as defined by the Stokes parameters, becomes a dynamical quantity: it is reduced in magnitude and it rotates.

The change in magnitude is handled by pseudorandomly generating pair-creation events that reduce the photon weight.
At each timestep of size $\Delta t$, the Stokes parameters of the photon are transformed to the basis defined by \cref{eq:LMAbasis} and the probability of pair creation $W^\pm(a, \eta_\gamma, S_j) \Delta \tau$ is evaluated.
An electron-positron pair is generated if $r_1 < W^\pm(a, \eta_\gamma, S_j) \Delta \tau$, where $r_1$ is a pseudorandomly generated number in the unit interval, $\Delta \tau = \Delta t / (\omega'/m)$, and $j = 1$ (LP) or $3$ (CP).
The pair creation rate is precalculated and tabulated as a function of $a$ and $\eta_\gamma$ for $S_j = \pm1$; the value for arbitrary $S_j$ is fixed by linear interpolation between the two extreme cases, because the rate is a linear function of the Stokes parameters.

If pair creation occurs, the harmonic index $n$ is obtained by inverting $r_2 = \text{cdf}(n)$, where $r_2$ is a pseudorandom number and $\text{cdf}(n) = \sum_{i = n^\star}^n W^\pm_i / W^\pm$ is the cumulative density function, precalculated and tabulated as a function of $a$, $\eta_\gamma$ and $n$ for $S_j = \pm1$.
The fraction of lightfront momentum transferred to the positron $s$ and azimuthal angle $\phi$ are obtained by rejection sampling of the double-differential rate, \cref{eq:PairCreationRate} (LP) or \cref{eq:CP_PairCreationRate} (CP).
The positron four-momentum $q'$ is constructed in the ZMF using
    \begin{align}
    q_\text{ZMF}^0 &= m \left(n \eta_\gamma / 2\right)^{1/2},
    &
    \abs{\vec{q}_\text{ZMF}'} &= m \left[n \eta_\gamma / 2 - (1 + \arms^2)\right]^{1/2},
    &
    \cos\theta_\text{ZMF} &= \frac{(1 - 2 s) q^0}{\abs{\vec{q}'}}
    \end{align}
and then transformed back to the laboratory frame.
The electron four-momentum $q$ follows from
    \begin{equation}
    k' + n k = q + q'.
    \end{equation}

As pair creation is much rarer than photon emission, \textit{Ptarmigan} incorporates a form of event biasing, where we artificially increase the pair-creation rate $W^\pm$ by a factor $R_\uparrow$, while reducing the weight of the daughter particles by the same factor: see Section 3.3 in \citet{blackburn.epjc.2022}.
Thus a photon with initial weight $w_\gamma$, creating an electron and positron with weights $w_\gamma / R_\uparrow$, becomes a photon with weight $w_\gamma (1 - 1 / R_\uparrow)$.
This accelerates the convergence of the statistical properties of pair-creation events, at the expense of generating many low-weight particles that must subsequently be tracked.

The photon polarization must rotate because it is a mixed state and the pair-creation rate is polarization-dependent.
Consider, for example, a photon with weight $w_\gamma$ and arbitrary $-1 \leq S_1 \leq 1$ (taking the background to be LP).
The weight in each eigenstate $S_1 = \pm 1$ is $w_\pm = w_\gamma (1 \pm S_1)/2$.
Decay to electron-positron pairs means that these individual weights are reduced to $w_\pm' = w_\gamma (1 \pm S_1) [1 - W^\pm(\arms, \eta_\gamma, \pm 1) \Delta \tau] / 2$ after a time interval $\Delta\tau$.
The decrease in the total weight $w_\gamma$ is therefore accompanied by a change in the Stokes parameters, as $S_1' = (w'_+ - w'_-) / (w'_+ + w'_-) \neq S_1$.
To account for this, the photon Stokes parameters are modified after each timestep to:
    \begin{align}
    S'_i =
        S_i \frac{1 - W^\pm(a, \eta_\gamma, 0) \Delta \tau}{1 - W^\pm(a, \eta_\gamma, S_j) \Delta \tau}
        - \delta_{ij} \frac{W^\pm(a, \eta_\gamma, 1) - W^\pm(a, \eta_\gamma, -1)}{2} \Delta \tau,
    \end{align}
where $\delta_{ij}$ is the Kronecker delta, $i \in \{1,2,3\}$,  and $j = 1$ (LP) or $3$ (CP).
This correction becomes significant only when pair creation itself becomes likely.\footnote{%
Equivalent logic applies to electron polarization, which changes due to the emission of radiation in a spin-dependent way.~\cite{tang.pra.2021}}
The Stokes parameters of the surviving photon $S'_i$ are then transformed back to the global basis (see \cref{fig:Basis}).

\subsection{Classical electrodynamics}
\label{sec:Classical}

It is possible to use the same simulation framework to model the interaction entirely classically.
In this case, the electron (positron) does not recoil at individual emission events, according to \cref{eq:PhotonRecoil}.
Instead, energy loss is accounted for by a radiation reaction force, here in the Landau-Lifshitz prescription~\cite{landau.lifshitz}.
Under the additional assumption that radiative losses are weak, i.e. that $q$ does not change significantly over a single cycle, we can define the following equation of motion for the quasimomentum $q$~\cite{smorenburg.lpb.2010}:
    \begin{equation}
    \frac{d q_\mu}{d \tau} = \frac{1}{2} m \partial_\mu a^2_\text{rms}(X) - \frac{2 \alpha}{3} m (\arms \eta_e)^2 q_\mu.
    \label{eq:LLEOM}
    \end{equation}
We solve \cref{eq:LLEOM} numerically to track charged particles as they travel through the EM field, emitting radiation.
This radiation is modelled by pseudorandomly generating `photons', even though these do not exist classically:
this works because the relevant physical observable in classical electrodynamics, the energy per unit frequency $\frac{d \mathcal{E}}{d \omega'}$, can be used to define a `number' spectrum via $\frac{d N}{d \omega'} = (\hbar \omega')^{-1} \frac{d \mathcal{E}}{d \omega'}$ (temporarily restoring factors of $\hbar$).
The emission algorithm is then adapted so that it uses the nonlinear Thomson spectrum, \cref{eq:LPClassicalPhotonRate}, rather than the nonlinear Compton spectrum, \cref{eq:LPPhotonRate}, as follows.

At each timestep, a photon is generated if a pseudorandom number $r_1 < P$, where $P = W^{\gamma,\text{cl}}(a, \eta_e) \Delta t / (q^0/m)$.
The total rate is precalculated by integrating \cref{eq:LPClassicalPhotonRate} over all $v$ and $\phi$, then summing over all $n$, and is tabulated as a function of $a$.
On emission, a harmonic index $n$ is obtained by inverting $r_2 = \text{cdf}(n)$, where the cumulative density function is itself tabulated as a function of $a$ and $n$.
We then sample $v$ and $\phi$ from \cref{eq:LPClassicalPhotonRate}, which define the photon momentum and polar scattering angle in the electron's instantaneous rest frame (IRF):
    \begin{align}
    \abs{\vec{k}_\text{IRF}'} &= \frac{m n \eta_e}{\sqrt{1 + \arms^2}},
    &
    \cos\theta_\text{IRF} &= 1 - 2 v.
    \end{align}
The four-momentum is then transformed back to the laboratory frame, using that the IRF travels at four-velocity $U = q / |q|$.
Pair creation has no classical equivalent and is therefore automatically disabled.

\subsection{LCFA-based dynamics}

The the locally constant field approximation (LCFA)~\cite{DiPiazza:2017raw,Ilderton:2018nws,DiPiazza:2018bfu,Seipt:2020diz} is the basis of the standard method by which strong-field QED processes are included in numerical simulations~\cite{ridgers.jcp.2014,gonoskov.pre.2015}.
The essential difference to the LMA is that photon emission and pair creation are assumed to occur instantaneously, i.e. that their formation lengths are much smaller than the laser wavelength.
The rates are therefore functions of the locally defined quantum parameter $\chi_{e,\gamma}$, which is defined in terms of the instantaneous (kinetic) momentum $\pi^\mu$;
the particle worldline ($x^\mu$) is defined at all, arbitrarily small, timescales.
In an LCFA-based simulation, charged particle trajectories follow from the Lorentz force equation:
    \begin{align}
    \frac{d \pi_\mu}{d \tau} &= \frac{q F_{\mu \nu} \pi^\nu}{m}
    &
    \frac{d x_\mu}{d \tau} &= \frac{\pi_\mu}{m},
    \end{align}
where $F_{\mu\nu}$ is the electromagnetic field tensor and $q = \pm e$ is the charge.
The electric field of a linearly polarized laser pulse is given by $\vec{E} = \text{Re}(\tilde{\vec{E}}) g(\varphi) + \text{Im}(\tilde{\vec{E}}) \frac{d g}{d \varphi}$, where $\tilde{\vec{E}}$ is the paraxial solution for the complex electric field of a focussed Gaussian beam~\cite{salamin.apb.2007} and $g(\varphi)$ is the pulse envelope: we include terms up to fourth-order in the diffraction angle $\vartheta = w_0 / z_R$.
A circularly polarized pulse is defined by adding together two linearly polarized pulses that have a phase difference of $\pi/2$ and orthogonal polarization vectors.

The trajectory is partitioned into timesteps of size $\Delta t$; photon emission occurs in a given timestep if a pseudorandom number $r < W^\gamma(\chi_e) \Delta t / (\pi^0/m)$, where $W^\gamma$ is the LCFA photon emission rate.
The total rate is precalculated using \cref{eq:LcfaPhoton} and stored as a lookup table in $\chi_e$.
When emission occurs, the photon momentum $\vec{k}'$ is constructed by first sampling the energy from the single-differential rate [\cref{eq:LcfaPhotonSingleDiff}], and then sampling the two angles from the triple-differential spectrum [\cref{eq:LcfaPhotonTripleDiff}].
The photon is then assigned a set of Stokes parameters using \cref{eq:LcfaStokes}.
The electron momentum after the scattering is fixed by conserving three-momentum, $\vec{\pi}' = \vec{\pi} - \vec{k}'$, which leads to a small, $O(1/\gamma^2)$, error in energy conservation~\cite{duclous.ppcf.2011}.
Pair creation is modelled in an analogous way, by tracking the photons along ballistic trajectories and sampling the LCFA pair-creation rate [\cref{eq:LcfaPair}].
An additional subtlety is that the Stokes parameters entering the rates are defined with respect to a basis that depends on the instantaneous electric and magnetic fields (see \cref{sec:CCF});
thus the Stokes parameters must be transformed at every timestep to match.

We can define a \emph{classical} LCFA framework in much the same way that we defined a classical LMA, by accounting for continuous radiative energy losses in the equations of motion themselves.
Charged particle trajectories are therefore obtained by numerically solving the Landau-Lifshitz equation~\cite{tamburini.njp.2010,vranic.cpc.2016}.
Photons are pseudorandomly generated along these trajectories by sampling the classical emission spectrum [\cref{eq:LCFAClassicalRate}].
We have also implemented a phenomenologically motivated \emph{modified classical} model which incorporates quantum corrections to the emission spectrum, but does not include stochastic recoil~\cite{duclous.ppcf.2011,blackburn.prl.2014}.
In this case, the radiation-reaction force is reduced in magnitude by the Gaunt factor~\cite{gonoskov.rmp.2022}, $0 < \mathcal{G}(\chi_e) < 1$, and photons are sampled from the QED emission spectrum [\cref{eq:LcfaPhotonTripleDiff}].

\subsection{Applicability}
\label{sec:Validity}

\changed{
The LMA is built on the assumption that the laser amplitude and wavevector are slowly varying functions of space and time.
Investigations in this and previous work support LMA-based simulations being accurate in practice if $N$, the number of cycles equivalent to the pulse duration, and $w_0$, the laser focal spot size, satisfy $N \gtrsim 4$ and $w_0 \gtrsim 2 \lambda$ (see Supplementary Materal in Ref.~\citenum{blackburn.njp.2021} and also Ref.~\citenum{blackburn.epjc.2022}).
Generally, there is no condition on the size of $\arms$ or $\eta$.
However, if classical RR is modelled using the LMA, we have the additional requirement that the energy loss per cycle is relatively small (see \cref{sec:Classical}) and therefore that $\arms^2 \eta_e \lesssim 30$.

The LCFA requires that the formation lengths of all QED processes be much smaller than the laser wavelength, or equivalently, $\arms \gg 1$ and $\arms^2 / \eta \gg 1$.
Note that even if these are satisfied, photon spectra are only accurate above the first Compton edge, $s > 2 \eta / (1 + \arms^2 + 2 \eta)$.
As a rule of thumb, for reasonable electron energies (up to 10s of GeV), the LCFA begins to be reliable above $a_0 \gtrsim 5$.
Provided that $a_0$ is large enough, there are in principle no restrictions on the pulse duration or focusing strength.
Nevertheless, the high-order paraxial approximation that \emph{Ptarmigan} uses to generate the laser fields in LCFA mode requires that $w_0 \gtrsim 2 \lambda$.

In LCFA mode, \emph{Ptarmigan} works in a similar way to the particle-in-cell (PIC) codes that include strong-field QED processes, with the exception that the fields in \emph{Ptarmigan} are prescribed, not self-consistently evolved.
PIC codes that have been extended to include strong-field QED, as well as spin and polarization dependence, include EPOCH (see Supplemental Material in \citet{gong.prl.2021}), OSIRIS (see \citet{qian.arxiv.2023}) and YUNIC (see references in \citet{song.prl.2022}).
These are mainly used to simulate laser-beam or laser-matter interactions~\cite{gonoskov.rmp.2022}.
Beam-beam interactions can be simulated using the dedicated code CAIN~\cite{cain}, which also includes self-consistent field evolution and strong-field QED processes under the LCFA.
It further implements an equivalent of the LMA for the modelling of laser-beam interactions; however, if the laser is linearly polarized, its coverage is limited to nonlinear Compton scattering and to $a_0 < 3$ (unlike Ptarmigan, which has coverage up to $a_{0} < 20$ for both nonlinear Compton and nonlinear Breit-Wheeler).
Other codes that use an equivalent of the LMA to simulate laser-beam interactions are NI~\cite{bamber.prd.1999} and IPstrong~\cite{hartin}.

Under certain conditions, it is not necessary to make approximations like the LMA or LCFA.
For example, if multiple emission effects are negligible, one can integrate the nonlinear Compton cross section over the collision phase space directly~\cite{krafft.prab.2023}.
This captures subharmonic structure in the radiation spectrum, but not secondary events like further photon emission or pair creation.
In the classical regime, the radiation spectrum can be obtained directly from the Li{\'e}nard-Wiechert integrals: see, for example, RDTX~\cite{thomas.prstab.2010} and its predictions of classical RR in laser-beam interactions~\cite{thomas.prx.2012}.
}

\section{Benchmarking}
\label{sec:Benchmarking}
   
\begin{figure}
    \centering
    \includegraphics[width=\linewidth]{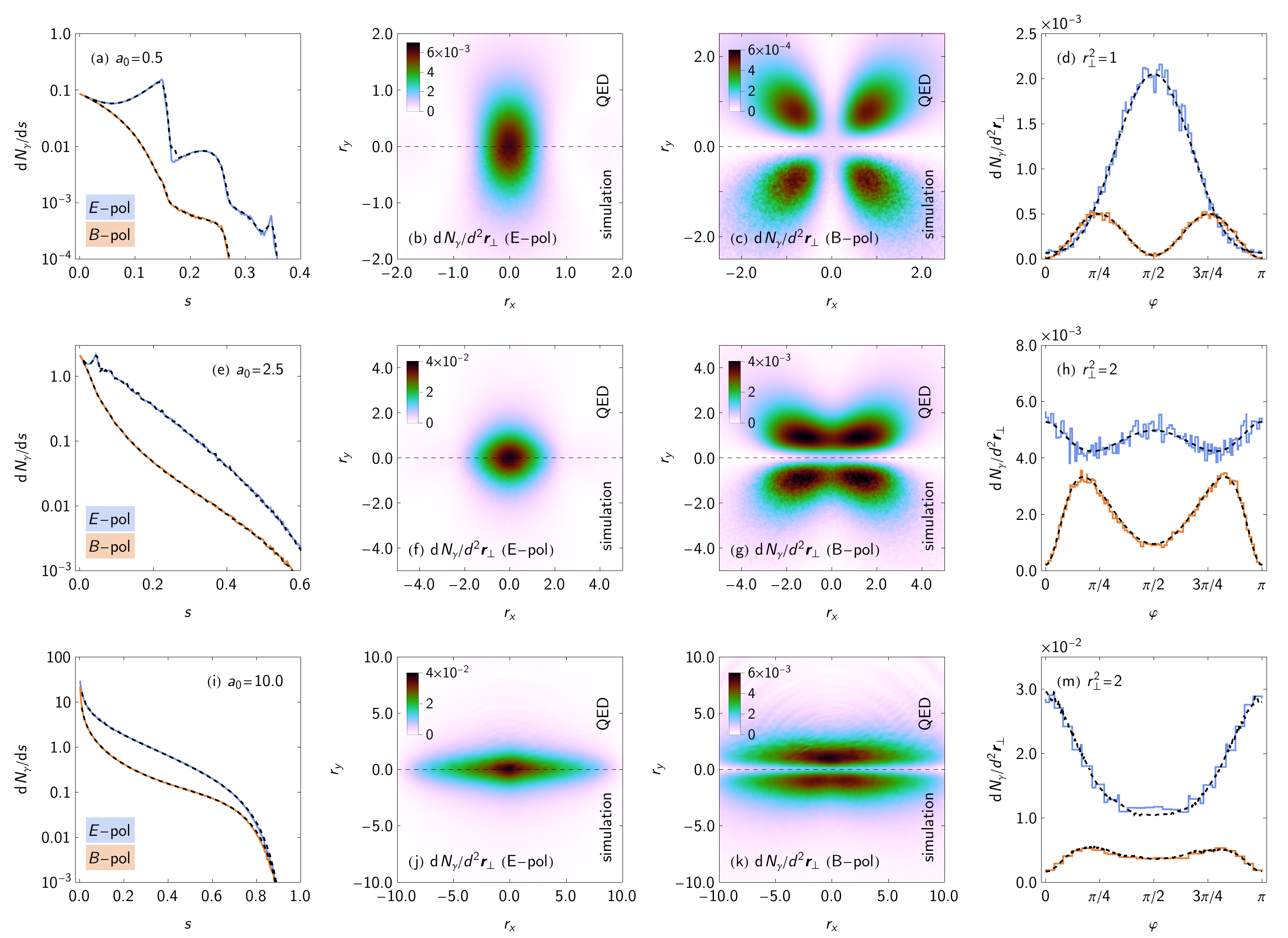}
    \caption{%
        Photon spectra predicted by LMA-based simulations and by QED, for electrons with energy parameter $\eta_e = 0.1$ colliding with linearly polarized, pulsed plane waves that have peak amplitude $a_0 = 0.5$ (top row), $2.5$ (middle row) and $10.0$ (bottom row):
        (first column) $d N_\gamma / d s$ from simulations (solid lines) and QED (black, dashed lines);
        (second and third columns) the polarization-resolved angular profiles of the emitted radiation;
        (fourth column) lineouts through the angular profiles at fixed $r_\perp$, from simulations (solid lines) and QED (black, dashed lines).
        }
    \label{fig:PhotonEmissionBenchmarks}
    \end{figure}

We begin by comparing the polarization-resolved spectra of photons emitted when a high-energy electron collides with an intense laser pulse, which we model as a 1D pulsed plane wave.
Its vector potential $e A^\mu(\varphi) = m a_0 \sin\varphi \, g(\varphi) \epsilon_1^\mu$, where $\epsilon_1^\mu = (0, 1, 0, 0)$ and the temporal envelope, $g(\varphi) = \cos^2[\varphi/(2N)]$, is non-zero for phases $\varphi$ that satisfy $\abs{\varphi} < N \pi$;
the number of cycles corresponding to the total duration of the pulse $N = 16$.
(The full-width-at-half-maximum duration of the intensity profile is $T\,[\text{fs}] \simeq 0.97 N \lambda\,[0.8~\micron] \simeq 15.5$.)
We choose three values of $a_0 \in \{0.5, 2.5, 10\}$ to illustrate the transition from the perturbative to nonperturbative regimes.
The energy parameter of the electrons is fixed at $\eta_e = 0.1$, which corresponds to an energy of 8.4~GeV for a head-on collision with a laser of wavelength of $0.8~\micron$.
Furthermore, in order to make comparisons with theory calculations that are first-order (single-vertex) in nature, we assume that this energy parameter does not change during the interaction with the pulse, i.e. we neglect quantum radiation reaction effects~\cite{blackburn.rmpp.2020}.
Our simulation and theory results are shown in \cref{fig:PhotonEmissionBenchmarks}.

Harmonic structure, which is clearly visible for $a_0 = 0.5$, is washed out as the laser intensity increases.
The first Compton edge is redshifted to smaller energies and the spectrum becomes increasingly synchrotron-like as the number of contributing harmonics increases.
The radiation is mainly polarized along the laser electric field, though the exact polarization purity is also photon-energy dependent: at very small $s$, $E$- and $B$-polarized photons are equally likely, whereas at the Compton edges, $E$-polarization dominates.
The angular profile of the $E$-polarized radiation changes from a dipole, at small $a_0$, to an ellipse elongated along the laser electric field, at large $a_0$.
The same compression in the $B$-direction may be seen in the angular profile of the $B$-polarized radiation, which changes from a quadrupole to a double ellipse.
The presence of an extinction line along $r_y = 0$ may be understood classically: for observers in this plane, which is also the plane of the trajectory, there is no vertical component of the electron current.
Our results demonstrate that simulations accurately reproduce what is expected from theory, both in terms of the absolute numbers and shapes of the spectra.
However, note that simulations have a finite cutoff for the largest harmonic order, so the high-energy tail of the spectrum will be underestimated.
Close examination of the first Compton edge, particularly in \cref{fig:PhotonEmissionBenchmarks}(a), reveals that the theory prediction is somewhat smoother than the simulation result.
This softening occurs because of interference at the scale of the pulse duration~\cite{King:2020hsk,Tang:2021qht}, which the LMA neglects:
the longer the pulse, the less significant this becomes.

\begin{figure}
    \centering
    \includegraphics[width=0.8\linewidth]{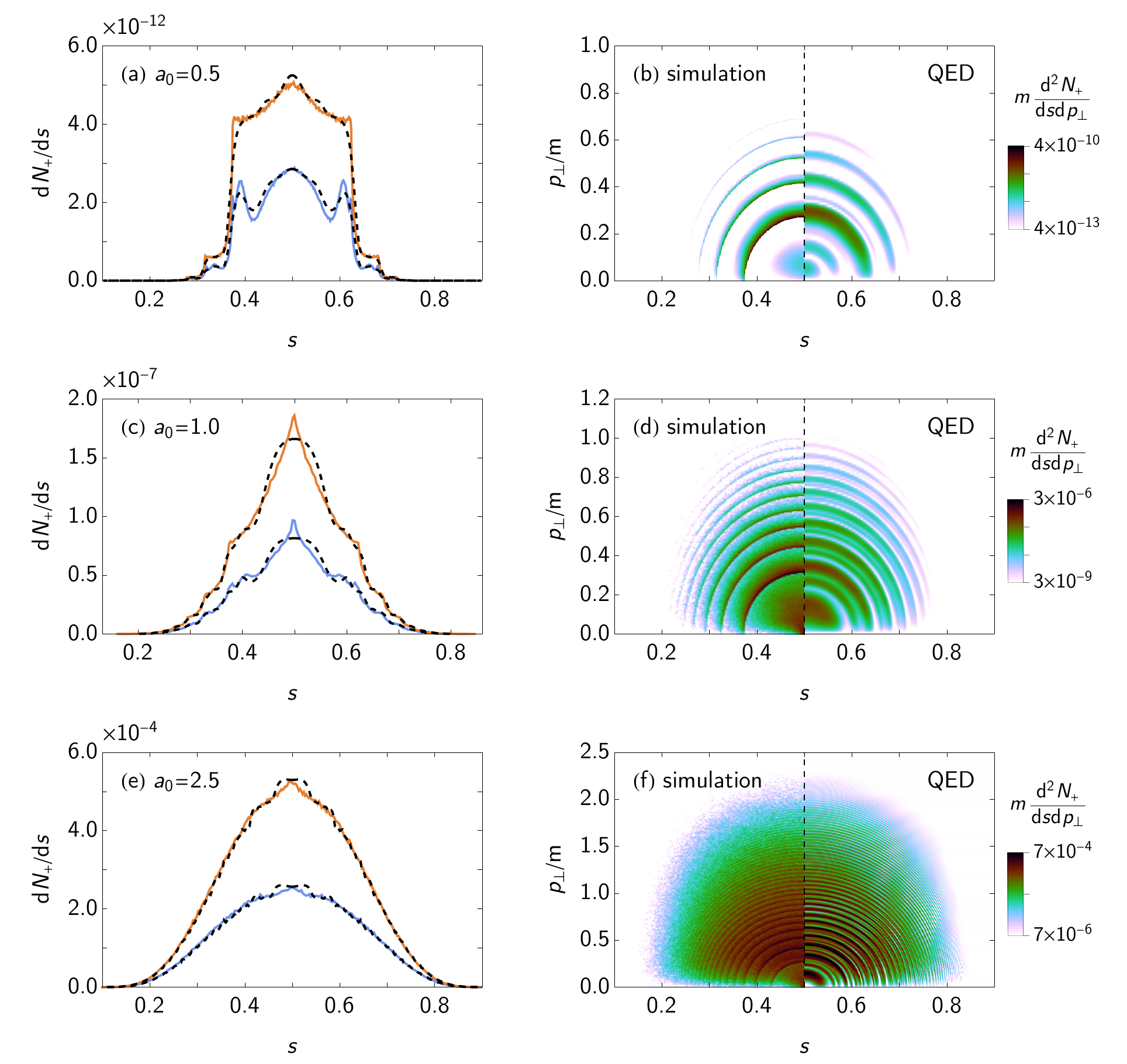}
    \caption{%
        Positron spectra for high-energy photons with energy parameter $\eta_\gamma = 0.2$ colliding with linearly polarized, pulsed plane waves that have peak amplitude $a_0$, wavelength $0.8~\micron$, and Gaussian temporal envelope.
        Left panels give the single-differential spectra for photons that are linearly polarized parallel (blue) or perpendicular (orange) to the laser electric field, as predicted by LMA-based simulations (solid lines) and by QED (dashed lines).
        Right panels give the log-scaled double-differential spectra for unpolarized photons.
        The $a_0$s have been chosen to illustrate the transition from the perturbative (multiphoton) regime in (a, b) to the quasistatic (tunnelling) regime in (e, f).
        }
    \label{fig:PairCreationBenchmarks}
    \end{figure}

We now consider positron production by high-energy photons colliding with an intense laser pulse.
In this comparison the laser temporal envelope is Gaussian, $g(\varphi) = \exp[-\varphi^2 / (4 N^2)]$, and $N = 16$.
(The full-width-at-half-maximum duration of the intensity profile is $T\,[\text{fs}] \simeq N \lambda\,[0.8~\micron] \simeq 16.0$.)
We choose three values of $a_0 \in \{ 0.5, 1.0, 2.5 \}$ to illustrate the transition from the  perturbative to quasistatic (tunnelling) regimes.
The energy parameter of the photons is fixed at $\eta_\gamma = 0.2$, which corresponds to an energy of 16.8~GeV for a laser wavelength of $0.8~\micron$.
The pair-creation probability is much smaller than one for all cases considered, so we use a rate biasing factor of $R_\uparrow \in \{ 10^{15}, 2 \times 10^{8}, 10^{5} \}$ to be able to resolve the positron spectrum.
Our simulation and theory results are shown in \cref{fig:PairCreationBenchmarks}.

As is the case in photon emission, the harmonic structure that is visible in the multiphoton regime, $a_0 \lesssim 1$, is washed out as $a_0$ increases.
The theory prediction is generally smoother than the simulation results because of pulse-envelope effects, which are more significant for a threshold process like pair creation.
The pulse contains a (small) range of frequency components and therefore, at fixed $s$, there is a range of threshold harmonic orders, spread around the LMA-predicted threshold order, $n^\star$.
This may be seen in the double-differential spectra, where the simulation results have clearly defined harmonics [observe the rings in the left-hand side of \cref{fig:PairCreationBenchmarks}(b)] and the theory results contain substructure between these harmonics.
Nevertheless, there is generally good agreement between the theory and simulations: notice that the pair creation probability increases by five orders of magnitude between $a_0 =0.5$ and $a_0 = 1.0$.
   
Finally, we provide a benchmark for \textit{Ptarmigan}'s classical electrodynamics mode.
We compare the simulation results against a direct calculation of the $s$-weighted spectrum, given the classical electron trajectory for two cases:
i) where radiation reaction is ignored, so the trajectory satisfies the Lorentz force equation;
and ii) where radiation reaction is accounted for, so the trajectory satisfies the Landau-Lifshitz equation.
In the latter case, the electron's energy parameter decreases as it propagates through the laser pulse.
Solving the Landau-Lifshitz equation for a circularly polarized, pulsed plane wave with envelope $g(\varphi)$, we find:~\cite{dipiazza.lmp.2008}
    \begin{align}
    \eta(\varphi) &= \frac{\eta_e}{1 + \frac{2}{3}\alpha a_0^2 \eta_e \mathcal{I}(\varphi)},
    &
    \mathcal{I}(\varphi) &= \int_{-\infty}^\varphi \left[ g^2(\psi) + \left( \frac{d g}{d\psi} \right)^2 \right] \, \rmd\psi,
    \end{align}
where $\eta_e$ is the initial energy parameter.
Using the LMA equations of motion [\cref{eq:LLEOM}] for a plane wave would give the same result, except that the derivative term in $\mathcal{I}(\varphi)$ would be absent; this is because LMA locally approximates variations in the envelope~\cite{heinzl.pra.2020}.

    \begin{figure}
    \centering
    \includegraphics[width=0.8\linewidth]{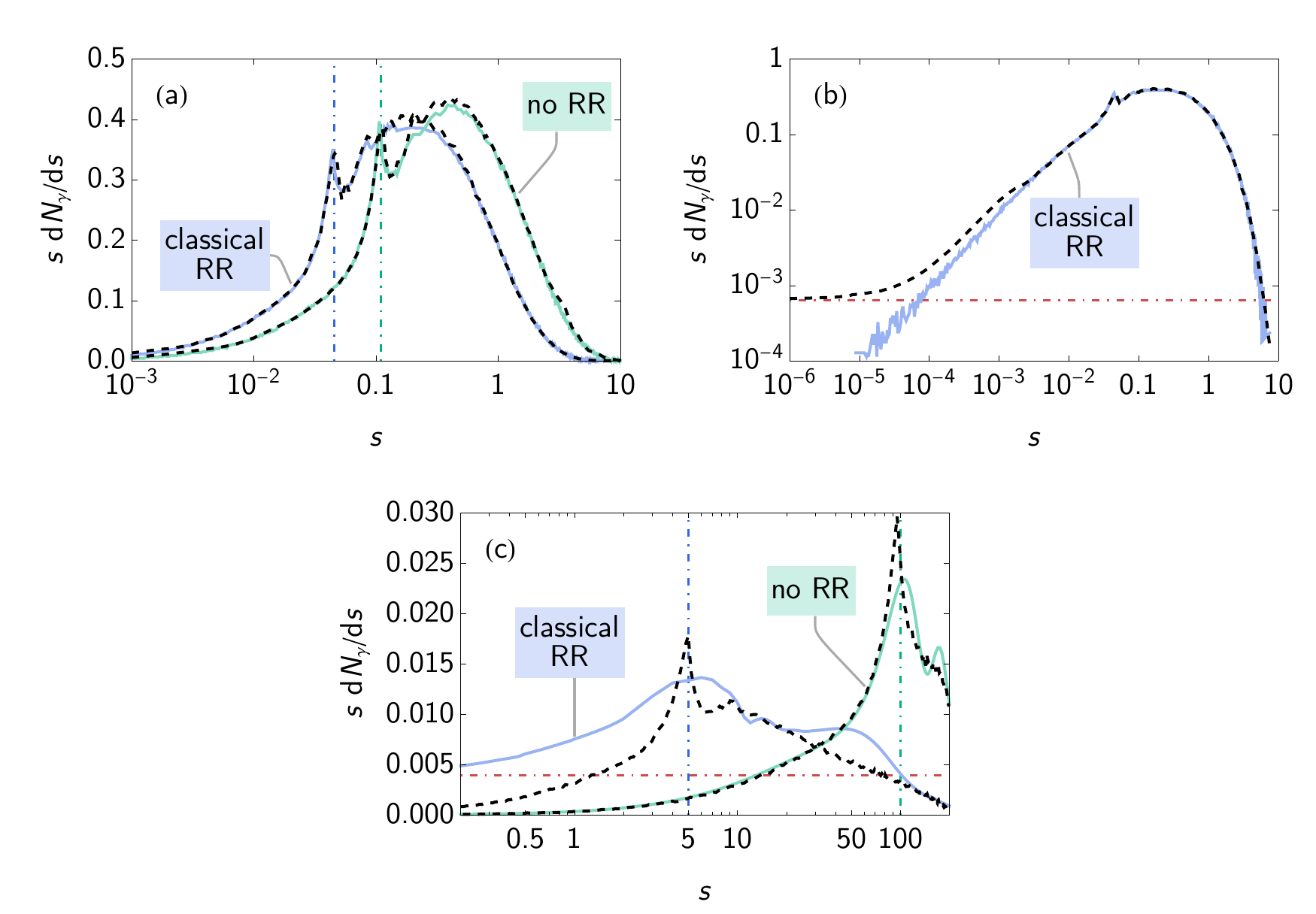}
    \caption{%
        \changed{(a, b)} Radiation spectra predicted by LMA-based simulations (solid lines) and classical theory (dashed lines), for electrons with energy parameter $\eta_e = 0.4$ colliding with circularly polarized, pulsed plane waves that have peak amplitude $a_0 = 2.5$, wavelength $0.8~\micron$, and duration equivalent to $N = 32$.
        \changed{(c) As in (a) and (b), but for electrons with $\eta_e = 100$ colliding with CP pulsed plane waves with $a_0 = 1.0$ and $N = 4$.}
        Vertical, dot-dashed lines in (a) and (c) give the positions of the first nonlinear Thomson edges predicted by \cref{eq:CRRedge}.
        The horizontal, dot-dashed lines in (b) and (c) gives \cref{eq:CRRlimit}, the IR limit expected from theory.
        There is no cutoff in classical electrodynamics so the spectrum extends beyond $s = 1$.
        }
    \label{fig:ClassicalBenchmarks}
    \end{figure}

Let us consider the case that $a_0 = 2.5$, $\eta_e = 0.4$ and $g(\varphi) = \cos^2[\varphi/(2N)]$, where $N = 32$.
(The full-width-at-half-maximum duration of the intensity profile is $31.0$~fs.)
We find that the final energy parameter $\eta_e' = \eta_e / (1 + \frac{1}{2} \pi \alpha a_0^2 \eta_e N) \simeq 0.209$, i.e. that the electron loses almost half its energy.
An energy loss of this magnitude manifests itself in a significant redshift of the first nonlinear Thomson edge, which is located at:
    \begin{equation}
    s_\text{edge} = \min_\varphi \frac{2 \eta^2(\varphi)}{\eta_e \left[1 + a_0^2 g^2(\varphi)\right]}.
    \label{eq:CRRedge}
    \end{equation}
We obtain $s_\text{edge} \simeq 0.0454$ with radiation reaction and $s_\text{edge} = 2 \eta_e / (1 + a_0^2) \simeq 0.110$ without.
Both are in good agreement with the results of LMA-based simulations and direct calculations of the Lienard-Wiechert integrals from classical electrodynamics: see \cref{fig:ClassicalBenchmarks}(a).
The photon spectrum, in the presence and absence of classical radiation reaction, is reproduced very well by the simulations.

There is, however, a discrepancy at very small $s$, shown in \cref{fig:ClassicalBenchmarks}(b).
The LMA predicts that $d N_\gamma / d s$ tends to a constant in the infrared limit,\footnote{The LCFA spectrum, by contrast, contains an integrable singularity in this limit~\cite{nikishov.jetp.1964}: $\lim_{s \to 0} (d N_\gamma/d s) \propto s^{-1/3}$.} i.e., that $\lim_{s \to 0} s \, (d N_\gamma / d s) \propto s$, whether radiation reaction is present or absent.
Naturally, the simulations obtain the same result.
However, it can be shown by regularising the plane wave result~\cite{PhysRevD.86.085037} that in the exact plane wave result, $dN_{\gamma}/ds$ diverges as $s\to0$, and~\cite{king.2023}:
    \begin{align}
    \lim_{s \to 0} s \frac{d N_\gamma}{d s} &=
       \frac{e^{2}}{4\pi^{2}} \left[ \left(\frac{2 + \Delta}{\Delta}\right) \ln(1 + \Delta) - 2 \right],
    &
    \Delta &= \lim_{\varphi\to\infty} \left[ \frac{\eta_{e}}{\eta(\varphi)} \right]^{2} - 1.
    \label{eq:CRRlimit}
    \end{align}
In the limit $\Delta\to 0$, the exact plane-wave result for the IR limit becomes $s\,(dN_{\gamma}/ds) \to (e^2/24\pi^2)\Delta^{2}$.
We note that logarithmic form of \cref{eq:CRRlimit} is similar in structure to the low-$\omega'$ limit of the energy spectrum $d \mathcal{E} /d\omega'$ derived by \citet{dipiazza.plb.2018}.
The non-zero IR limit for the energy spectrum originates from the fact the electron moves with reduced velocity after the collision; as it is associated with timescales much longer the laser period, the discrepancy is found at very small $s$, where envelope effects are important and the LMA is not accurate.

\changed{Finally, we present a more extreme example, in which LMA-based simulations are expected to fail.
We set the electron energy parameter to $\eta_e = 100$ and the laser amplitude and duration to $a_0 = 1.0$ and $N = 4$.
In this case $a_0^2 \eta_e = 100$, which means that the electron loses a significant fraction of its energy in a single cycle and therefore we cannot assume that its quasimomentum is slowly varying (see \cref{sec:Validity}).
We see from \cref{fig:ClassicalBenchmarks}(c) that, while the simulations are accurate in the no-RR case, they reproduce only the gross structure and redshifting of the spectrum when classical RR is included.
Interference effects not captured by the LMA mean that a distinct Compton edge, expected to be located at $s \simeq 5$ according to \cref{eq:CRRedge}, does not emerge;
the broad spectral feature appearing at $s \simeq 50$ is completely missed for the same reasons.
}

\section{Examples}
\label{sec:Examples}

Here we present two examples of the physics that can be explored with polarization-resolved simulations.

\subsection{Trident pair creation}

Electron-laser collisions can produce a large flux of photons with energies comparable to that of the incident electron.
The probability that these photons create pairs, and therefore fail to escape the pulse, depends not only the photon's momentum but also its polarization.
At large $a_0$ and small $\chi_\gamma$, for example, $B$-polarized photons are twice as likely to pair create as $E$-polarized photons.
The fact that electron-laser collisions produce mainly $E$-polarized photons means that the positron yield is overestimated by simulations that use spin-averaged and summed probability rates~\cite{king.pra.2013}.
As \textit{Ptarmigan} incorporates polarization dependence in both photon emission and pair creation, we consider how the yield of positrons changes when the polarization of the intermediate photon is taken into account, as an example.

    \begin{figure}
    \centering
    \includegraphics[width=0.8\linewidth]{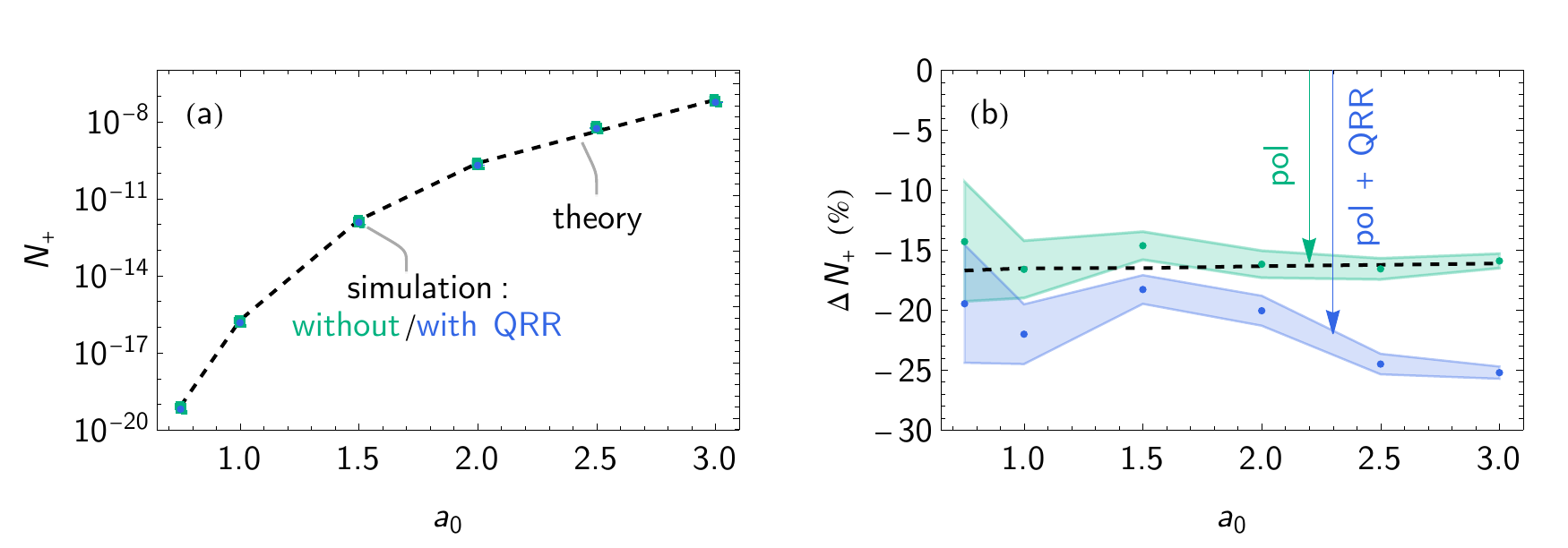}
    \caption{%
        (a) Positron yield per incident electron for electrons with energy $E_e = 16.5$~GeV ($\eta_e \simeq 0.197$) colliding with linearly polarized laser pulses with peak amplitude $a_0$ and duration equivalent to $N = 16$ cycles: simulation results, accounting for photon-polarization effects, with (blue points) and without (green points) quantum radiation reaction.
        (b) The change in the positron yield when taking into account: the polarization of the intermediate photon (green points) and additionally radiative energy losses (blue points).
        The coloured bands indicate the uncertainty in the simulation results at the single standard deviation level.
        Our results are crosschecked against theory data from \citet{tang.prd.2023} (black, dashed lines).
        }
    \label{fig:TridentBenchmarks}
    \end{figure}

The rapid growth of the pair-creation probability with increasing photon energy means that the dominant contribution to the trident positron yield comes from the tail of the photon spectrum.
Resolving this tail with Monte Carlo simulations requires large statistics: the results presented in \cref{fig:TridentBenchmarks} are the mean and standard deviation obtained from an ensemble of $N_c$ simulated collisions, where each collision includes $10^7$ primary electrons and $N_c = \{200, 100, 20, 5, 5, 5\}$ for $a_0 = \{0.75, 1.0, 1.5, 2.0, 2.5, 3.0\}$ respectively.
To resolve the pair creation itself we set the rate biasing factor to $R_\uparrow = \{10^{21}, 10^{17}, 10^{13}, 10^{11}, 10^{10}, 10^{8}\}$, respectively.
We set the electron energy parameter to $\eta_e = 0.197$, which is equivalent to an energy of 16.5~GeV for a head-on collision, the laser pulse envelope to $g(\varphi) = \cos^2[\varphi/(2N)]$, where $N = 16$, and vary $a_0$ between 0.75 and 3.0.
In order to compare our results with a direct numerical calculation of the two-step trident yield~\cite{tang.prd.2023}, we initially disable the recoil associated with photon emission, i.e. quantum radiation reaction.

Simulations show that taking the polarization of the intermediate photon into account reduces the yield by ${\sim}16\%$ across the full range of $a_0$ we have considered, which is consistent with the theoretical result.
The yields themselves are consistent with the theory results at the 2\% level (for $a_0 \geq 1$) and 5\% level (for $a_0 = 0.75$), albeit that simulations predict fewer positrons than expected.
This is because the nonlinear Compton rate is only summed up to a finite cutoff harmonic $n_\text{max}$.
When quantum radiation reaction is enabled, the positron yield is further reduced.
This is because radiative energy losses reduce the electron energy parameter $\eta_e$ and therefore the energy parameters of the photons that go on to produce electron-positron pairs.
This correction becomes increasingly important as $a_0$ rises.

\subsection{Polarization-resolved angular profiles}

Measuring the angular profile of the radiation emitted in an electron-laser collision has been proposed as means of inferring the laser amplitude $a_0$, because the profile effectively carries information about the electron transverse momentum~\cite{harshemesh.ol.2012,blackburn.prab.2020}.
Consider a high-energy electron colliding head-on with a circularly polarized laser pulse with envelope $g(\varphi)$.
The transverse momentum as a function of phase is $p_\perp(\varphi) = m a_0 g(\varphi)$;
assuming that the longitudinal momentum is sufficiently large that the Lorentz factor $\gamma \gg a_0^2$, the angle between the electron momentum and the collision axis is $\theta(\varphi) \simeq a_0 g(\varphi) / \gamma$.
As the radiation is strongly beamed along the instantaneous momentum, the angular size of the profile $\sim a_0 / \gamma$.

    \begin{figure}
    \centering
    \includegraphics[width=0.7\linewidth]{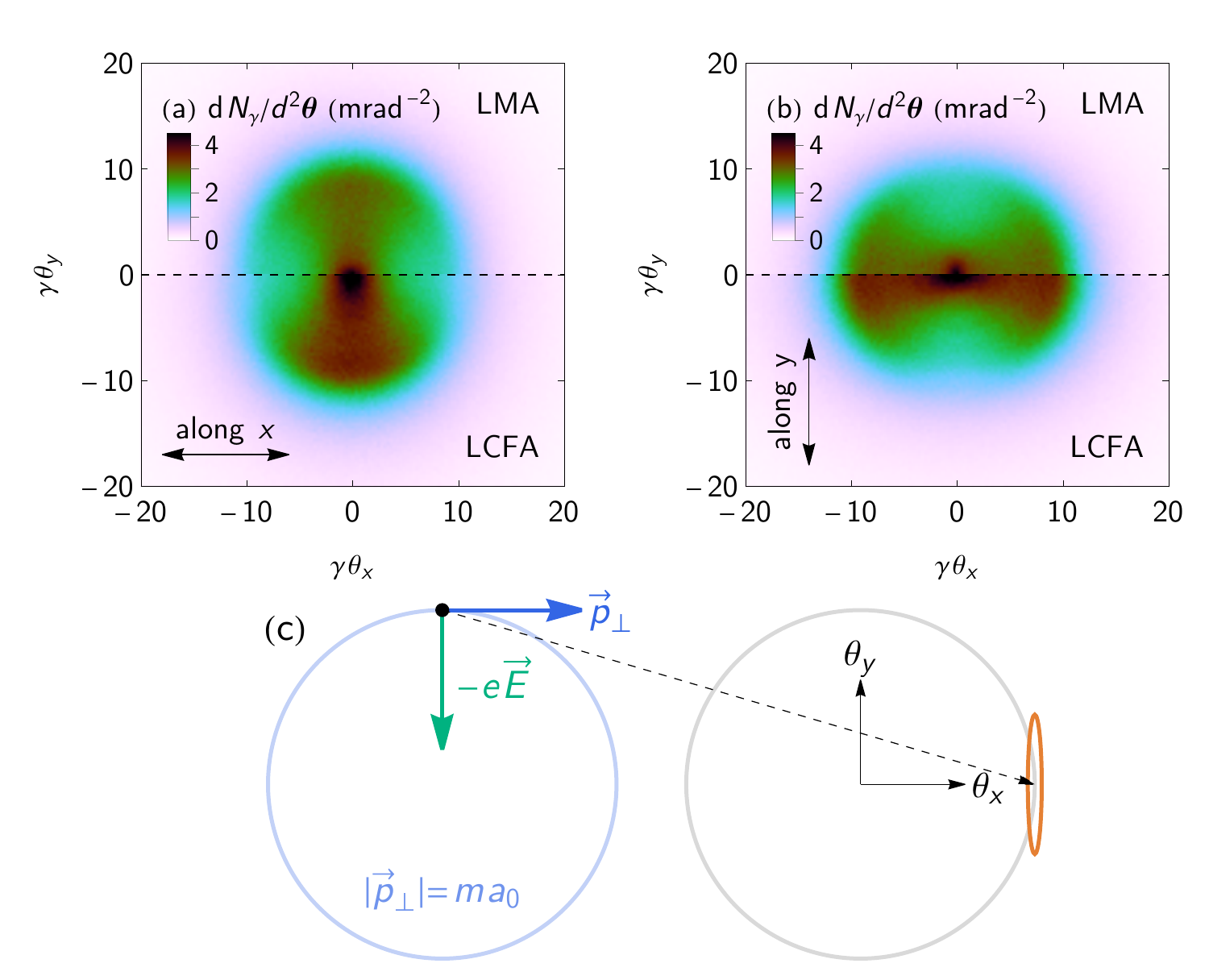}
    \caption{%
        Angular profile of the radiation emitted when an energy with energy parameter $\eta_e = 0.2$ collides with a circularly polarized laser pulse of amplitude $a_0 = 10$, as predicted by LMA- and LCFA-based simulations:
        (a) horizontally and (b) vertically polarized components.
        The origin of the angular structure is illustrated in (c): the electron is accelerated by the laser electric field (in green) on a circular orbit (in blue), emitting radiation (in orange) that is mostly polarized along the instantaneous acceleration.
        }
    \label{fig:BatSignal}
    \end{figure}

Let us consider the angular profile of the radiation emitted when an electron with initial energy parameter $\eta_e = 0.2$ collides with a circularly polarized laser pulse of amplitude $a_0 = 10$ (the pulse envelope $g(\varphi) = \cos^2[\varphi/(2N)]$ and $N = 16$), assuming that we also resolve the polarization of emitted radiation.
\Cref{fig:BatSignal} gives the angular distributions expected if we select photons that are linearly polarized along the horizontal or vertical axes.
We see that a clear ``batwing'' structure emerges, with extinction regions lined up along the polarization axis.
This relationship may be understood classically with the help of the diagram in \cref{fig:BatSignal}(c), which shows the electron trajectory in a monochromatic, circularly polarized plane wave, as viewed along the collision axis (or laser wavevector).
The crucial point is that in the transverse plane the electron's instantaneous momentum, $\vec{p}_\perp$, and acceleration, $-e \vec{E}/m$, are perpendicular to each other.
In a constant, crossed field, radiation is polarized along the direction of the instantaneous acceleration.
Thus a photon, emitted by a electron with $p_x = m a_0$ and $p_y = 0$ (as shown in the diagram), travels horizontally (i.e. in $x$) and is polarized vertically (i.e. in $y$).
Selecting the horizontally polarized component, for example, then leads to an extinction region along the horizontal axis: see \cref{fig:BatSignal}(a).

The same structure emerges in both LMA- and LCFA-based simulations because $a_0 = 10$ and the photon formation length is small compared to the laser wavelength.
In our previous work~\cite{blackburn.njp.2021}, we observed that the LMA effectively moves the fast oscillation of the trajectory into the QED rates.
The same phenomenon occurs here: the structure in the polarization-resolved angular profile comes from the azimuthal angle dependence in the Stokes parameters, \cref{eq:CPNLCStokes}, rather than the trajectory itself.
We conclude that, in the same way that the polarization-summed angular profile contains information about the transverse momentum, the polarization-resolved angular profile additionally contains information about the transverse acceleration.

\section{Summary}

We have presented a simulation framework based on the locally monochromatic approximation (LMA), which enables us to predict strong-field QED interactions in the perturbative ($a_0 \ll 1$), transition ($0.5 \lesssim a_0 \lesssim 2$), and nonperturbative regimes ($a_0 \gg 1$).
The limitations of this approach are that i) the background field must be sufficiently `plane-wave-like', which sets bounds on the duration of the laser pulse and ii) the computational cost of evaluating very high-order harmonics restricts our implementation of the LMA to normalized amplitudes $a_0 \leq 20$.
Nevertheless, this goes considerably further than any previous code and ensures good overlap with the region in which the LCFA is accurate.
In contrast to our previous work~\cite{blackburn.njp.2021,blackburn.epjc.2022}, where we examined circularly polarized lasers and unpolarized $\gamma$ rays, we have considered here the physically richer problem of linearly polarized lasers.
The broken symmetry of this case makes the numerical implementation more challenging, because the loss of azimuthal symmetry means that an additional integral must be performed to evaluate the probability rates.
It also makes it necessary to account for $\gamma$-ray polarization effects, because the radiation emitted by an electron in a linearly polarized background is preferentially polarized along the direction of the laser electric field.

The open-source \textit{Ptarmigan} code~\cite{ptarmigan} can now be used to simulate strong-field QED interactions: in linearly or circularly polarized, plane-wave or focussed, laser pulses; using QED, classical or modified-classical models of the particle dynamics; with either LMA- or LCFA-based probability rates.
Fine-grained control of the physics under consideration can be achieved by enabling (or disabling) radiation reaction, electron-positron pair creation, or the polarization dependence thereof.
Our benchmarking against theoretical calculations of nonlinear Compton scattering, nonlinear Breit-Wheeler pair creation, and trident pair creation shows that the code achieves per cent level accuracy across the whole transition regime ($0.5 \lesssim a_0 \lesssim 2$).
This accuracy is also maintained at higher $a_0$, where the LMA automatically recovers the LCFA where it should do so~\cite{heinzl.pra.2020}.
The \textit{Ptarmigan} simulation framework is designed to be extensible and additional physics can be included, as motivated by experimental needs.
Future work will include the role of fermion spin and higher order processes, such as vacuum polarization.

\begin{acknowledgments}
T.G.B. thanks Kyle Fleck (Queen's University Belfast) for contributions to the \textit{Ptarmigan} source code.
The computations were enabled by resources provided by the Swedish National Infrastructure for Computing (SNIC) at the High Performance Computing Centre North (HPC2N), partially funded by the Swedish Research Council through grant agreement no. 2018-05973.
\end{acknowledgments}

\section*{Data availability}
The \textit{Ptarmigan} source code can be obtained from its Github repository~\cite{ptarmigan}.
The version used in this work (v1.3.1) is archived at \href{https://doi.org/10.5281/zenodo.7957000}{https://doi.org/10.5281/zenodo.7957000} (Ref.~\citenum{ptarmigan-zenodo}).

\bibliography{references}

\appendix

\section{Circularly polarized plane waves}
\label{sec:CPwave}

The laser polarization is defined by the Stokes parameter $S_3^\text{laser} = \pm 1$.
$S_3^\text{laser} = 1$ denotes \emph{right}-circular polarization, which means that the electric field rotates anticlockwise around the direction of propagation, or equivalently that the laser photons have positive helicity (right handedness).
$S_3^\text{laser} = -1$ denotes left-circular polarization or that the laser photons have negative helicity (left handedness): this is the default setting for circularly polarized lasers in \textit{Ptarmigan}.

\subsection{Photon emission}

The single-differential emission rate per unit proper time, at a particular harmonic index $n$, in the LMA is given by~\cite{heinzl.pra.2020}
    \begin{equation}
    \frac{\rmd W^\gamma_n}{\rmd s} =
        -\alpha m
        \left\{
            J_n^2(z) +
            \frac{\arms^2}{2}
            \left[ 1 + \frac{s^2}{2(1-s)} \right]
            \left[ 2 J_n^2(z) - J_{n-1}^2(z) - J_{n+1}^2(z) \right]
        \right\}
    \label{eq:CP_EmissionRate}
    \end{equation}
where the argument of the Bessel functions $z$ and auxiliary variables are
    \begin{align}
    z^2 &= \frac{4 n^2 \arms^2}{1 + \arms^2}
        \frac{s}{s_n (1-s)}
        \left[ 1 - \frac{s}{s_n (1-s)} \right],
    &
    s_n &= \frac{2 n \eta}{1 + \arms^2}
    \end{align}
and $0 < s < s_n / (1 + s_n)$.
The azimuthal angle $\phi$ is uniformly distributed in $[0, 2\pi)$.

The Stokes parameters of the emitted photon are given by~\cite{ivanov.epjc.2004}
    \begin{equation}
    \vec{S} = \begin{pmatrix}
        -{\cos 2\phi} & -{\sin 2\phi} & 0 \\
        \sin 2\phi & -{\cos 2\phi} & 0 \\
        0 & 0 & 1
    \end{pmatrix}
    \begin{pmatrix}
        S'_1 \\
        S'_2 \\
        S'_3 
    \end{pmatrix}
    \label{eq:CPNLCStokes}
    \end{equation}
where
    \begin{align}
    S'_1 &= \frac{2}{S'_0} \left\{ \left[ J_{n-1}^2(z) + J_{n+1}^2(z) - 2 J_n^2(z) \right] + 4 \left( 1 - \frac{n^2}{z^2} + \frac{1}{2 \arms^2} \right) J^2_n(z) \right\}, \\
    S'_2 &= 0, \\
    S'_3 &= \frac{S_3^\text{laser}}{S'_0} \left( 1 - s + \frac{1}{1-s} \right) \left[ 1 - \frac{2 s}{s_n (1-s)} \right] \left[ J^2_{n-1}(z) - J^2_{n+1}(z) \right]
    \end{align}
and
    \begin{equation}
    S'_0 = \left( 1 - s + \frac{1}{1-s} \right) \left[ J_{n-1}^2(z) + J_{n+1}^2(z) - 2 J_n^2(z) \right] - \frac{4}{\arms^2} J_n^2(z).
    \end{equation}
The rotation matrix ensures that the Stokes parameters are defined with respect to the basis given in \cref{eq:LMAbasis}.
Note that this is the only place that the azimuthal angle $\phi$ appears explicitly.

\paragraph{Classical limit.}

In the limit that $\eta \ll 1$ (for arbitrary $a$), the partial rates take the form
    \begin{align}
    \frac{\rmd W^{\gamma,\text{cl}}_n}{\rmd v} &=
        \frac{\alpha m n \eta}{1 + \arms^2}
        [\arms^2 J_{n-1}^2(z) + \arms^2 J_{n+1}^2(z) - 2(1+\arms^2) J_n^2(z)],
    &
    z^2 &= \frac{4 \arms^2 n^2 v (1-v)}{1 + \arms^2},
    \label{eq:LowEtaRate}
    \end{align}
where $0 \leq v = s/s_n \leq 1$.
The Stokes parameters are obtained by replacing $1 - s + 1/(1-s) \to 2$ and $1 - 2 s / [s_n (1-s)] \to 1 - 2v$.

\subsection{Pair creation}

The double-differential pair-creation rate per unit time, $W^\pm$, at a particular harmonic index $n$, is given by~\cite{tang.prd.2022} (see also \citet{nikishov.jetp.1964})
    \begin{multline}
    \frac{\rmd^2 W^\pm_n}{\rmd s \rmd \phi} =
        \frac{\alpha m}{2\pi}
        \left\{
            J_n^2 +
            \frac{\arms^2}{4}
            \frac{s^2 + (1-s)^2}{s (1-s)}
            \left( J_{n-1}^2 + J_{n+1}^2 - 2 J_n^2 \right)
    \right. \\
            {} - (S_1 \cos2\phi + S_2 \sin2\phi) \left[
                \left( \frac{2 n^2 \arms^2}{z^2} - 1 - \arms^2 \right) J_n^2
                - \frac{\arms^2}{2} \left( J_{n-1}^2 + J_{n+1}^2 \right)
            \right]
    \\ \left.
            {} - S_3^\text{laser} S_3 \left[
                \frac{\arms^2}{4}
                \frac{s^2 + (1-s)^2}{s (1-s)}
                \left( 1 - \frac{2}{s_n s (1-s)} \right)
                \left( J_{n-1}^2 - J_{n+1}^2 \right)
            \right]
        \right\},
    \label{eq:CP_PairCreationRate}
    \end{multline}
where
    \begin{align}
    z^2 &= \frac{4 n^2 \arms^2}{1 + \arms^2}
        \frac{1}{s_n s(1-s)}
        \left[ 1 - \frac{1}{s_n s (1-s)} \right],
    &
    s_n &= \frac{2 n \eta}{1 + \arms^2},
    \end{align}
and
    \begin{equation}
    \frac{1}{2} \left[ 1 - \sqrt{1 - 4/s_n} \right]
    < s <
    \frac{1}{2} \left[ 1 + \sqrt{1 - 4/s_n} \right].
    \end{equation}
The second term, which depends on $S_1$ and $S_2$, disappears on integration over the azimuthal angle $\phi$.
Pair creation is likelier for photons with $S_3 = S_3^\text{laser}$, i.e. that have the same helicity as the background, than for photons with $S_3 = -S_3^\text{laser}$.
The total rate is obtained by integrating \cref{eq:CP_PairCreationRate} over all $s$ and $\phi$, then summing over all harmonic orders $n \geq n^\star = \lceil 2(1+\arms^2)/\eta \rceil$.
It depends on the normalised amplitude $\arms$, energy parameter $\eta_\gamma$ and only the third Stokes parameter $S_3$, i.e. $W_\pm = W_\pm(a, \eta_\gamma, S_3)$.

\section{Constant crossed fields}
\label{sec:CCF}

Rates calculated for constant, crossed fields form the basis of the locally constant, crossed fields approximation, which has become the standard method by which strong-field QED processes are included in numerical simulations~\cite{ridgers.jcp.2014,gonoskov.pre.2015}.

\subsection{Photon emission}

The double-differential emission rate per unit proper time is given by~\cite{baier.1998}
    \begin{equation}
    \frac{\rmd^2 W^\gamma}{\rmd u \rmd \zeta} =
        \frac{2 \alpha m}{3 \sqrt{3} \pi \chi_e}
        \frac{u}{(1+u)^3}
        \left\{
            \zeta^{2/3} [1 + (1+u)^2] - (1+u)
        \right\}
        K_{1/3}\!\left( \frac{2 u \zeta}{3 \chi_e} \right),
    \label{eq:LcfaPhotonTripleDiff}
    \end{equation}
where $u = \omega' / (\gamma m - \omega')$ and $\zeta = [2\gamma^2(1-|\vec{v}|\cos\theta)]^{3/2}$.
Here $\theta$ is the polar angle in the \emph{lab} frame, measured with respect to the electron's instantaneous velocity $\vec{v}$, and $\gamma$ is the electron Lorentz factor.
The domain of \cref{eq:LcfaPhotonTripleDiff} is $0 \leq u < \infty$ and $1 \leq \zeta < \infty$; the azimuthal angle is uniformly distributed in $[0,2\pi)$.
A useful approximation for ultrarelativistic particles is $\zeta \simeq (1 + \gamma^2\theta^2)^{3/2}$ as $\theta \sim O(1/\gamma)$.

The polarization of the emitted photon is fixed with respect to the orthonormal basis $\vec{e}_1 = \vec{a} - (\vec{n}\cdot\vec{a}) \vec{n}$ and $\vec{e}_2 = \vec{a} \times \vec{n}$, where $\vec{a}$ is the unit vector along the electron's instantaneous acceleration $\vec{E} + \vec{n}\times\vec{B}$ and $\vec{n}$ is the unit vector along the photon's 3-momentum.
Let $\beta$ be the angle between $\vec{n}$ and the plane defined by the electron's instantaneous velocity and acceleration: $\beta = \vec{n} \cdot (\vec{v} \times \vec{a}) \sim O(1/\gamma)$.
Then the Stokes parameters of the emitted photon are~\cite{baier.1998}:
    \begin{align}
    \begin{split}
    S_1 &= \frac{1}{S_0} \left[ \mu^2 K^2_{2/3}(\nu) - \beta^2 K^2_{1/3}(\nu) \right]
    \\
    S_2 &= 0
    \\
    S_3 &= \frac{2}{S_0} \left[ 1 + \frac{u^2}{2(1+u)} \right] \beta \mu K_{1/3}(\nu) K_{2/3}(\nu)
    \end{split}
    \label{eq:LcfaStokes}
    \end{align}
where
    \begin{equation}
    S_0 = \mu^2 K^2_{2/3}(\nu) + \beta^2 K^2_{1/3}(\nu) + \frac{u^2 \mu^2}{2(1 + u)} \left[ K^2_{1/3}(\nu) + K^2_{2/3}(\nu) \right]
    \label{eq:LcfaStokesZero}
    \end{equation}
and $\mu^2 = \beta^2 + 1/\gamma^2$ and $\nu = u \gamma^3 \mu^3 / (3 \chi_e)$.
As $S_1^2 + S_2^2 + S_3^2 < 1$, the photon is only partially polarized.
Integrating over all $\zeta$, we obtain
    \begin{equation}
    \frac{\rmd W^\gamma}{\rmd u} =
        \frac{\alpha m}{\sqrt{3} \pi}
        \frac{1}{(1+u)^3}
        \left\{
            [1 + (1+u)^2] K_{2/3}(\xi)
            - (1+u) \int_\xi^\infty \! K_{1/3}(t) \, \rmd t
        \right\}.
    \label{eq:LcfaPhotonSingleDiff}
    \end{equation}
where $\xi = 2 u / (3 \chi_e)$.
Finally we obtain the total rate by integrating over all $u$:
    \begin{align}
    W^\gamma &= \frac{\sqrt{3}\alpha m}{2\pi} \chi_e \mathcal{H}(\chi_e),
    &
    \mathcal{H}(\chi_e) &= \frac{2}{9\chi_e} \int_0^\infty \frac{5 u^2 + 7 u + 5}{(1+u)^3} \, K_{2/3}\!\left( \frac{2 u}{3\chi_e} \right) \, du
    \label{eq:LcfaPhoton}
    \end{align}

\paragraph{Classical limit.}
The classical limit of \cref{eq:LcfaPhotonTripleDiff} is given by:
    \begin{equation}
    \frac{\rmd^2 W^{\gamma,\text{cl}}}{\rmd u \rmd \zeta} =
        \frac{2 \alpha m}{3 \sqrt{3} \pi \chi_e}
        u
        \left(
            2 \zeta^{2/3} - 1
        \right)
        K_{1/3}\!\left( \frac{2 u \zeta}{3 \chi_e} \right).
    \label{eq:LCFAClassicalRate}
    \end{equation}
As there is no cutoff in photon frequency, $u = \omega' / (\gamma m)$.
The Stokes parameters are obtained by setting $u^2/(1+u) \to 0$ in \cref{eq:LcfaStokes,eq:LcfaStokesZero}; as a result, $S_1^2 + S_2^2 + S_3^2 = 1$.
The classical total rate is obtained by setting $\mathcal{H}(\chi_e) = 5\pi/3$ in \cref{eq:LcfaPhoton}.
    
\subsection{Pair creation}

The pair creation probability rate is used to determine the positron's energy (as a fraction $f$ of the photon energy $\omega'$), polar angle $\theta$ and azimuthal angle $\phi$ (as defined with respect to $\vec{n}$, the unit vector along the photon's 3-momentum).
The rate is a function of the photon's quantum parameter $\chi_\gamma$ and its polarization.
We define the latter in terms of the three Stokes parameters $S_{1,2,3}$ and the orthonormal basis $\vec{e}_1 = \vec{a} - (\vec{n}\cdot\vec{a}) \vec{n}$, $\vec{e}_2 = \vec{a} \times \vec{n}$, where $\vec{a}$ is the unit vector along the instantaneous acceleration $\vec{E} + \vec{n}\times\vec{B}$.
The triple-differential rate per unit time (resolved in energy and polar and azimuthal angle) is given by
    \begin{multline}
    \frac{\rmd^3 W^\pm}{\rmd f \rmd \zeta \rmd \phi} =
        \frac{\alpha m}{2 \sqrt{3} \pi^2}
        \left\{
            \left[ 1 + \zeta^{2/3} \frac{f^2 + (1-f)^2}{f(1-f)} \right] K_{1/3}(\delta \zeta)
        \right. \\
        + S_1 \left[ \cos 2\phi - \zeta^{2/3} ( 1 + \cos 2\phi) \right] K_{1/3}(\delta \zeta)
        \\
        - S_2 \sin 2\phi \, (\zeta^{2/3} - 1) K_{1/3}(\delta \zeta)
        \\ \left.
        + S_3 \sin\phi \left[ \zeta^{1/3} (\zeta^{2/3} - 1) \frac{f^2 + (1-f)^2}{f(1-f)} \right] K_{2/3}(\delta \zeta)
        \right\},
    \end{multline}
where $\zeta = \left[ 2 \gamma^2 (1 - |\vec{v}| \cos\theta )\right]^{3/2} \geq 1$ is a transformed polar angle that depends on the positron Lorentz factor $\gamma$ and velocity $\vec{v}$ and the auxiliary variable
    \begin{equation}
    \delta = \frac{2}{3 \chi_\gamma f (1-f)}.
    \end{equation}
The $S_2$- and $S_3$-dependent terms vanish on integration over the azimuthal angle~\cite{baier.1998}:
    \begin{equation}
    \frac{\rmd^2 W^\pm}{\rmd f \rmd \zeta} =
        \frac{\alpha m}{\sqrt{3}\pi}
        \delta
        \left[
            1 + \zeta^{2/3} \left( \frac{f}{1-f} + \frac{1-f}{f} - S_1 \right)
        \right]
        K_{1/3}(\delta \zeta).
    \label{eq:LcfaPair}
    \end{equation}
Photon-polarization dependence appears in the form of the Stokes parameter $S_1$, which gives the degree of linear polarization with respect to $\vec{e}_1$ and $\vec{e}_2$.
The sign indicates that pair creation is more probable for photons polarized perpendicular to the acceleration, $S_1 = -1$, than for photons polarized parallel to the acceleration, $S_1 = 1$.
As expected, the spectrum is symmetric around $f = 1/2$.

\end{document}